\documentclass[fleqn,usenatbib]{mnras}

\usepackage[utf8]{inputenc}
\usepackage{threeparttable}
\usepackage{amsmath}
\usepackage{graphicx}
\usepackage{epstopdf}
\usepackage{threeparttable}
\usepackage{graphicx, color}
\usepackage{bm}
\usepackage{soul}
\usepackage{amssymb}
\usepackage{hyperref}
\usepackage{gensymb}

\newcommand{\eqb}{\begin{eqnarray}}
\newcommand{\eqe}{\end{eqnarray}}

\title[Radiative Signatures of Reconnection]
{Radiative Signatures of Plasmoid-Dominated Reconnection in Blazar Jets}
\author[]{I.M.~Christie$^1$\thanks{E-mail: ichristi231@gmail.com} , M.~Petropoulou$^2$\thanks{E-mail: m.petropoulou@astro.princeton.edu} , L.~Sironi$^{3}$, \& D.~Giannios$^1$ \\
$^1$Department of Physics, Purdue University, 525 Northwestern Avenue, West Lafayette, IN, 47907, USA \\
$^2$Department of Astrophysical Sciences, Princeton University, 4 Ivy Lane, Princeton, NJ 08544, USA \\
$^3$Department of Astronomy, Columbia University, 550 W 120th Street, New York, NY 10027, USA
}
 
\begin{document}

\date{Received.../Accepted...}

\pagerange{\pageref{firstpage}--\pageref{lastpage}} \pubyear{2017}

\maketitle

\label{firstpage}

\begin{abstract}
The multi-wavelength spectral and temporal variability observed in blazars set tight constraints on current theoretical emission models. Here, we investigate the relativistic magnetic reconnection process as a source of blazar emission in which quasi-spherical plasmoids, containing relativistic particles and magnetic fields, are associated with the emission sites in blazar jets. 
By coupling recent two-dimensional particle-in-cell simulations of relativistic reconnection with a time-dependent radiative transfer code, we compute the non-thermal emission from a chain of plasmoids formed during a reconnection event. The derived photon spectra display characteristic features observed in both BL Lac sources and flat spectrum radio quasars, with the distinction made by 
varying the strength of the external photon fields, the jet magnetization, and the number of pairs per proton contained within. Light curves produced from reconnection events are composed of many fast and powerful flares that  appear on excess of a slower evolving envelope produced by the cumulative emission of medium-sized plasmoids.  The observed variability is highly dependent upon the orientation of the reconnection layer with respect to the blazar jet axis and to the observer. Our model provides a physically motivated framework for explaining the multi-timescale blazar variability across the entire electromagnetic spectrum. 
\end{abstract}


\section{Introduction}
\label{sec:intro}
Blazars are a subclass of active galactic nuclei (AGN) that were originally identified by their flat radio spectra, compact emitting regions, and variable and polarized emission at radio and optical wavelengths. Only within the last decade have blazars become ubiquitous sources of the extragalactic X-ray and $\gamma$-ray sky. Their emission does not only extend over the whole electromagnetic spectrum, but also varies on a range of timescales; i.e., from months \citep{ahnen2016} to minutes [e.g. PKS~2155-304 \citep{aharonian2007}, Markarian~501 \citep{albert2007}, 3C~279~\citep{ackerman2016}, 3C~54.3 \citep{britto2016}]. It is generally accepted that the broadband blazar emission originates from a relativistic jet emerging from an accreting supermassive black hole (SMBH), with the jet axis aligned with the observer's line of sight \citep{blandford_rees_1978,urry1995}. However, it is not yet clear whether the multi-timescale and multi-wavelength blazar variability is related to stochastic processes occurring in the accretion disk of the SMBH \citep{sobolewska2014} or in the relativistic jet.

%%%%%% FIGURE BEGIN %%%%%%%%
\begin{figure*}
\centering
\includegraphics[width=0.8\textwidth]{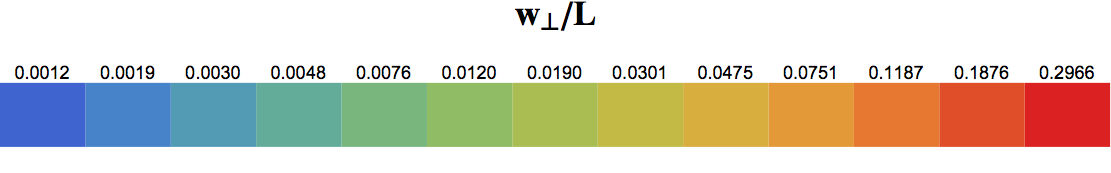}\\
    
\includegraphics[height=0.435\textwidth]{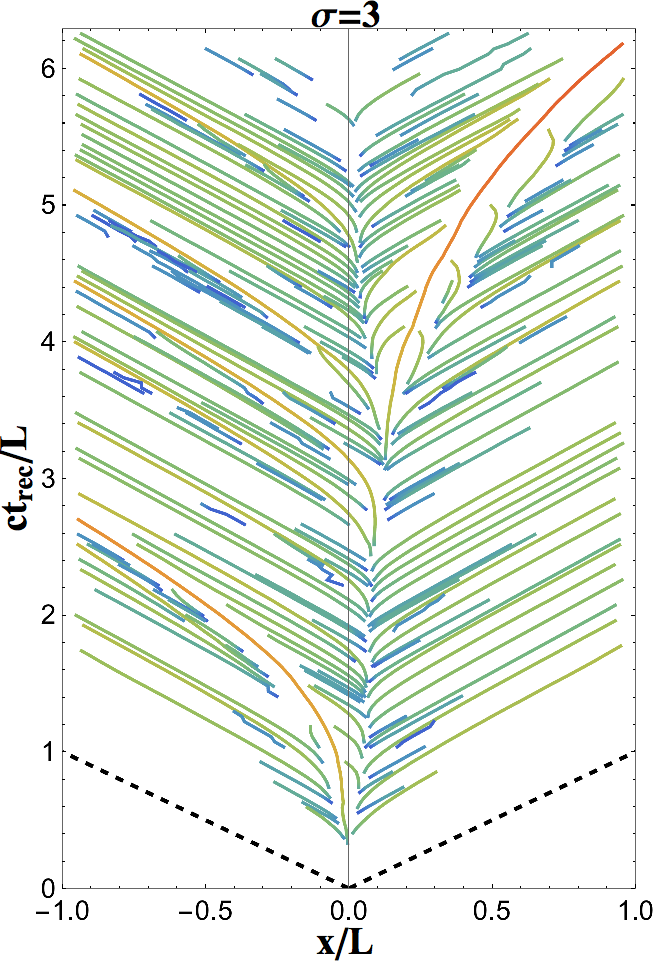}
%\hfill
\includegraphics[height=0.435\textwidth]{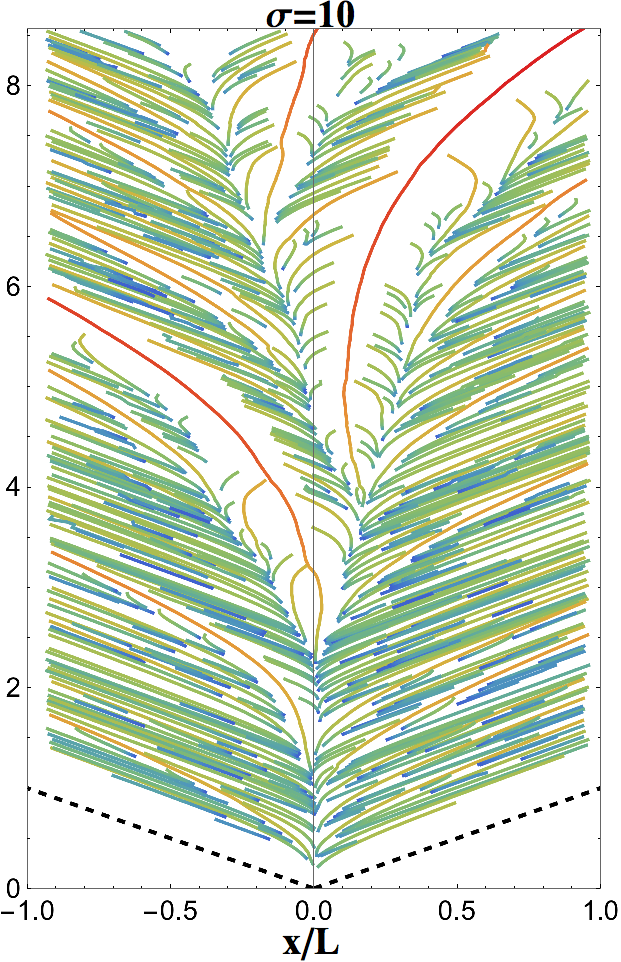}
%\hfill
\includegraphics[height=0.435\textwidth]{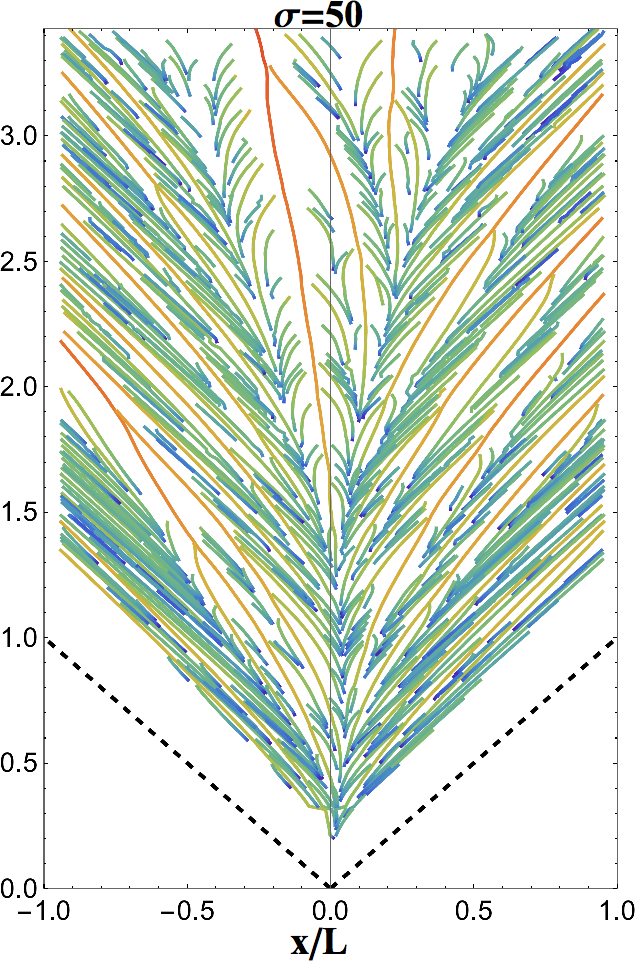}
\caption{Time-position plots of the plasmoids' centers as derived from 2D PIC simulations of pair plasma for three different values of the magnetization ($\sigma=3$, $10$, and $50$). Only plasmoids that pass our selection criterion (see Sec.~\ref{sec:plas_select}) are shown here. The time is measured in the reconnection frame (i.e. jet's co-moving frame) and is normalized to $L/c$, where $L$ is the half-length of the layer. The colour coding denotes the  transverse size $w_\perp$ (i.e., perpendicular to the reconnection layer) of an individual plasmoid, normalized to $L$. The black dashed lines denote objects moving at the speed of light. A coloured version of this plot is available online. Movies displaying the evolution of the reconnection layer's structure can be found at \url{https://goo.gl/XwkbQA}. }
\label{fig:tracks}
\end{figure*}
%%%%%% FIGURE END %%%%%%%%	
Quasi-spherical blobs, containing relativistic particles and magnetic fields, are often postulated to exist in the blazar jet in order to explain the variable, broadband blazar emission \citep{mastichiadis1995,bloom1996, chiaberge1999,celotti2007}. Although the physical origin of theses blobs is unknown, a strong candidate is the  magnetic reconnection process \citep{giannios2009, giannios13}. Instabilities occurring in the blazar jet can result in the production of current sheets where the reconnection process is triggered \citep{spruit2001,giannios2006,barniol2017, gill2018}. These current sheets are susceptible to tearing instabilities which fragment the sheets into a chain of magnetic islands or \textit{plasmoids} \citep{loureiro2007,uzdensky2010,fermo2010,huang2012,loureiro2012,takamoto2013}, each containing relativistic particles and magnetic fields. Plasmoids produced via reconnection are physically-motivated candidates for the blobs invoked in blazar emission models.

The dynamics and properties of the plasmoids formed by the reconnection process as well as the radiative, non-thermal processes operated in these regions can only be studied self-consistently by means of kinetic particle-in-cell (PIC) simulations. Such simulations have been performed, in both two and three dimensions (2D and 3D), in the relativistic regime, where the magnetic energy density exceeds the rest mass energy density of the plasma \citep{guo2014,sironi2014,guo2015,sgp16,werner2016,rowan2017,werner2018,ball2018}. \cite{spg15}, in particular, using 2D PIC simulations of reconnection in electron-positron and electron-proton plasmas, showed that the reconnection process satisfies the basic requirements needed for modeling blazar emission: i) efficient dissipation of magnetic energy into energy of accelerated particles, ii) an extended, non-thermal distribution of relativistic particles, and iii) plasmoids characterized by a rough equipartition between magnetic fields and relativistic particles.  The statistical properties of the {\sl plasmoid chain}, such as the plasmoid size and velocity distributions, can only be investigated if the numerical simulations extend to sufficiently long spatial and temporal scales, as demonstrated by \cite{sgp16} (hereafter, denoted as SGP16). The 2D PIC results of SGP16  were later incorporated by \cite{pgs16} (hereafter, denoted as PGS16) into a radiative model for the evolution of the radiating particles contained within an individual plasmoid. PGS16 provided a physically-motivated model for flares powered by individual plasmoids and derived approximate, analytical expressions for the peak luminosity and flux doubling timescale of a flare as a function of the plasmoid's size and momentum.

The aim of this study is to expand upon the work of PGS16 by determining the emission produced from the entire plasmoid chain in an application to the blazar multi-wavelength variability. 
To achieve this, we combine the 2D PIC results of SGP16, which describe the plasmoids' growth and evolution within the reconnection layer, with a radiative transfer model tracking the evolution of the particle distribution in each plasmoid and the resulting photon spectrum.
	
This paper is structured as follows: In Sec.~\ref{sec:pic_results} we briefly summarize the PIC results of SGP16 and the plasmoid properties adopted in our model. In Sec.~\ref{sec:model}, we introduce our method of computing the emission from the plasmoid chain. Our results for individual plasmoid-powered flares and the entire plasmoid chain are presented in Sec.~\ref{sec:results} while a discussion and summary is provided in Sec.~\ref{sec:discussion}.

\section{Summary of 2D PIC Results}
\label{sec:pic_results}
In SGP16, the authors employed a set of 2D PIC simulations of pair plasmas for three values of the plasma magnetization\footnote{The magnetization, or the ratio of the magnetic energy density over the rest mass energy density of the plasma, is defined as $\sigma = B_{\rm up}^{2} / 4 \pi m_{\rm p} n_{\rm up} c^{2}$, where $B_{\rm up}$ and $n_{\rm up}$ are the magnetic field strength and particle number density of the cold plasma far from the reconnection layer.} ($\sigma = 3$, $10$, and $50$). These simulations, which were initiated in a Harris-sheet configuration with no guide field, showed that the relativistic reconnection process can naturally produce a hierarchical chain of plasmoids. In this work, we will use the results from these simulations in determining the evolution of the photon and particle distributions for each plasmoid and the entire plasmoid chain. Henceforth, we will use \textit{particles}, \textit{electron-positron pairs}, and \textit{pairs} interchangeably. Below, we list the main assumptions regarding the adopted PIC results which enter our calculations for the emission produced by the plasmoid chain:
\begin{enumerate} 
\item The co-moving particle number density $n_{\rm co}$ and magnetic field strength $B$ within each plasmoid are assumed to be constant in time (see panels d-f and j-l in Fig. 5 of SGP16).

\item We use the area-averaged plasmoid properties as derived from the 2D PIC simulations and neglect their radial dependence from the plasmoid's center (see Appendix~A in SGP16). Although this is a simplifying assumption, it does not alter our main conclusions (see Sec.~\ref{sec:discussion}). 

\item In 2D simulations, the plasmoid is a 2D quasi-circular structure. The transverse diameter (i.e. perpendicular to the layer), denoted as $w_{\perp}$, is Lorentz invariant while the longitudinal diameter (i.e. parallel to the layer) of the plasmoid as measured in its co-moving frame is $\sim 3 w_{\perp}/2$ (see panels a-c of Fig.~5 in SGP16). In this work, we use 2D PIC results to construct a 3D model of the emitting region by considering the plasmoid as an ellipsoid\footnote{The properties of plasmoids in 3D (e.g. shape and statistics) are not yet robustly determined and they depend on the assumed guide field strength. We thus take the shape to be an ellipsoid for simplicity.}
% As such, a plasmoid's shape and properties is not well characterized in 3D and is dependent upon the assumed guide field. As such, we 
whose third dimension is taken to be $w_{\perp}$. The volume of a plasmoid, as measured in its co-moving frame, can therefore be estimated as $V(t) \approx \pi w_{\perp}(t)^{3} /4$.

\item We ignore the dynamical effects of radiative losses on the plasmoid's structure, as well as the effect the plasmoid's radiation has on its neighbors. This radiation, combined with the relative motion of a plasmoid with respect to its neighbors, can provide an increase in the observed emission from its neighbors (see Appendix~C of \citealt{pgs16}), while exerting an external Compton drag force \citep{beloborodov2017}. The role of these effects on a neighboring plasmoid's dynamics and observed emission will be presented elsewhere (Christie et al., in preparation).

\item Due to the long time-scales and length-scales of the employed PIC simulations, we were able to demonstrate the self-similar nature of the reconnection process (SGP16). The distribution of plasmoid sizes forms a power law, extending from a few plasma skin depths (i.e., the microscopic characteristic plasma scale) to $\sim 0.1$ of the layer's length (see Fig.~6 in SGP16). This holds irrespective of the ratio between the macroscopic layer length $L$ and the microscopic plasma scale $r_{\rm L}$ \citep[see also][]{pcsg17}. Additionally, the largest plasmoids contain the highest energy particles, whose Larmor radius is a few percent of the layer's length, once again irrespective of the ratio $L/r_{\rm L}$ (see Fig.~11 in SGP16). Based on these two findings, we can conclude that the reconnection process in pair plasmas is self-similar up to the astrophysically-relevant regime  of $L/r_{\rm L}\gg 1$.
\end{enumerate}

The adopted data of SGP16, which contain all information regarding a plasmoid's growth and dynamics, require a small amount of manipulation before running our radiative transfer calculations (for details, see  Appendix~\ref{sec:smoothing}). Below we discuss the selection of plasmoids from the three simulations with different magnetization %each $\sigma$ simulation 
and also present those properties of the plasmoid chain that are pertinent to our radiative transfer model. 

\subsection{Plasmoid Selection \& Properties}
\label{sec:plas_select}
%%%%%% FIGURE BEGIN %%%%%%%%
\begin{figure}
\centering
\includegraphics[height=0.3\textwidth]{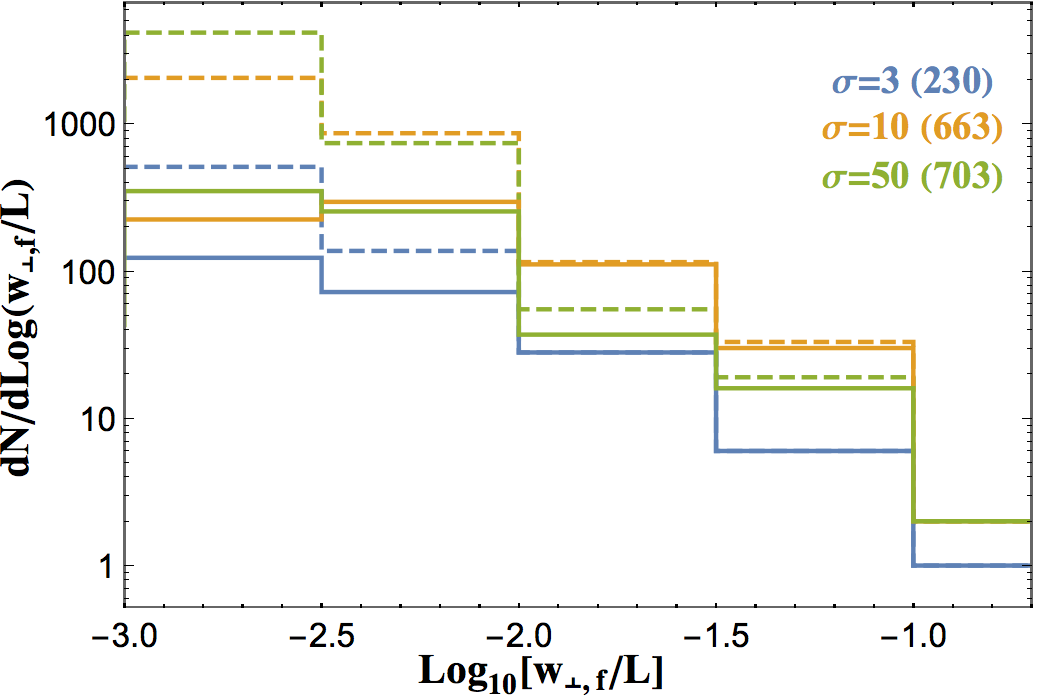}
\caption{Histogram displaying the distribution of the plasmoid's final transverse size, normalized to $L$, for the three magnetizations considered, as marked on the plot. The dashed lines show the final sizes of all plasmoids formed during the corresponding PIC simulation while the solid lines represent those plasmoids which pass our selection criterion (see Sec.~\ref{sec:plas_select}). The total number of plasmoids considered in the radiative transfer calculations are displayed in parenthesis next to their respective $\sigma$ value. A coloured version of this plot is available online.}
\label{fig:final_size_hist}
\end{figure}
%%%%%% FIGURE END %%%%%%%%
Plasmoids are born within the reconnection layer due to tearing instabilities occurring in the current sheet \citep{uzdensky2010}.
They may accelerate up to approximately the Alfv{\'e}n velocity $v_{\rm A}/c = \sqrt{\sigma/(1+\sigma)}$ \citep{lyubarsky2005}, grow via accretion from smaller plasmoids and from the unstructured outflow, coalesce with one another, and advect out of the layer. The plasmoid birth and merger rate increase for larger values of $\sigma$ (SGP16).

Many small plasmoids (i.e. $w_{\perp}/L \lesssim 10^{-3}$ where $L$ is the half-length of the reconnection layer) quickly merge with other neighboring plasmoids soon after they are born, thereby having short lifetimes. %of a fraction of a dynamical time $L/c$. 
These short-lived plasmoids are not expected to contribute significantly to the observed emission from the entire plasmoid chain, given that they contain a small number of particles \citep[see also Fig.~17 in][]{pcsg17}.
% little energy. \cite{pcsg17} have shown, using analytical estimates of individual flares, that flares powered by short-lived (i.e. small) plasmoids obtain luminosities much lower than the average luminosity of the entire reconnection event. 
We therefore exclude all plasmoids with lifetimes $\lesssim 0.08 \, L/c$, as measured in the reconnection frame, from our radiative transfer calculations. %We also neglect plasmoids that are still present within the reconnection layer at the end of the simulation, as we do not have information regarding their full evolution. This latter criteria, however, only removes $\sim 10$ plasmoids per simulation, regardless of $\sigma$. 
Under this selection criterion, we end up with $230$~plasmoids for $\sigma=3$, $663$~plasmoids for $\sigma=10$, and $703$~plasmoids for $\sigma=50$. 

%%%%%% FIGURE BEGIN %%%%%%%%
\begin{figure*}
\centering
\includegraphics[height=0.36\textwidth]{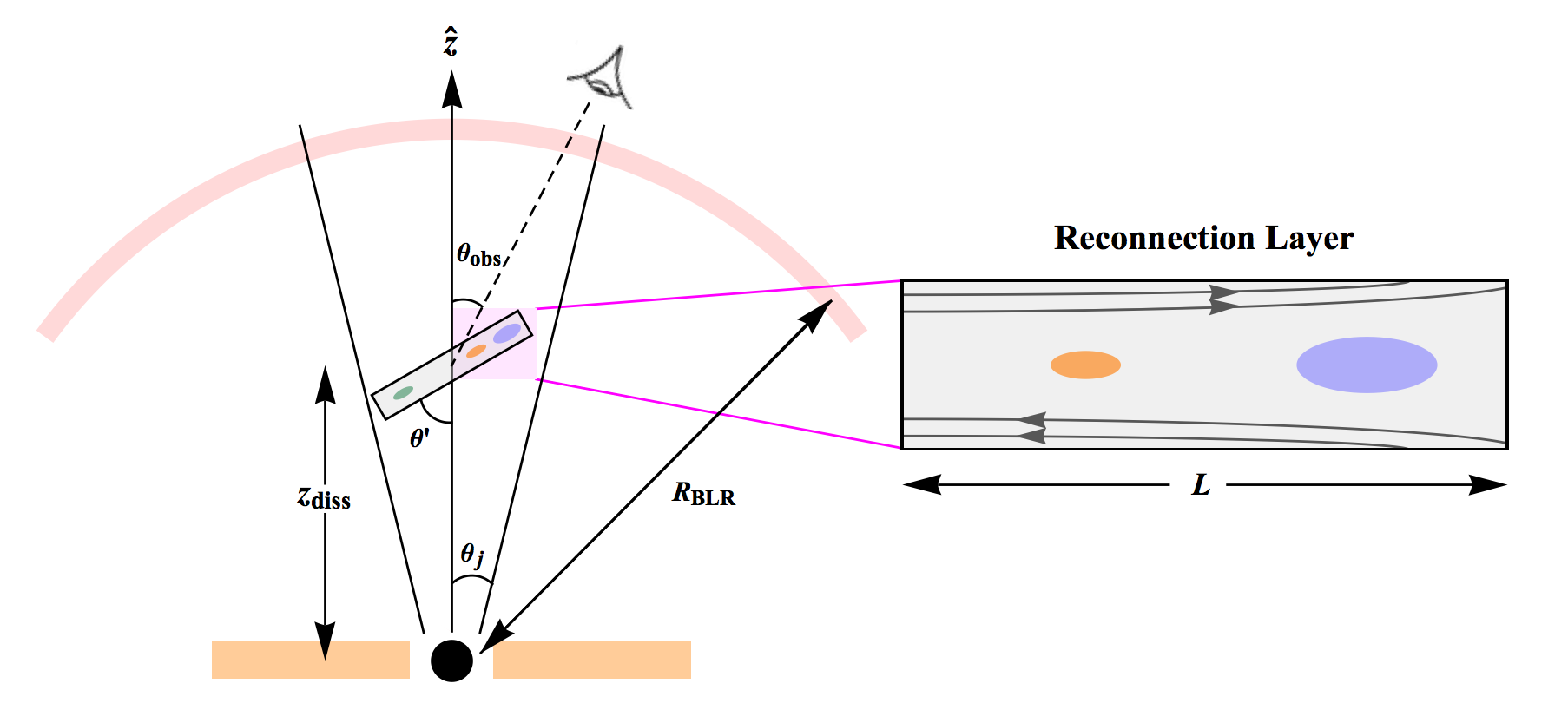}
\caption{Schematic diagram illustrating our model. The reconnection layer forms within the relativistic jet at a distance $z_{\rm diss}$ from the SMBH and is oriented at an angle $\theta^{\prime}$ with respect to the jet axis, as measured in the jet's rest frame. The inset plot on the right portrays plasmoids, whose dynamics are determined using PIC simulations, moving within the layer along with the reconnecting magnetic field lines. We also consider emission from the BLR, assumed to be a spherical shell located at a radial distance $R_{\rm BLR}$ from the SMBH. All results are mapped to an observer, positioned at an angle $\theta_{\rm obs}$ with respect to the jet axis. Note that objects in our schematic are not drawn to scale.} 
\label{fig:layer_in_jet_sketch}
\end{figure*}
%%%%%% FIGURE END %%%%%%%%

Figs.~\ref{fig:tracks} and~\ref{fig:final_size_hist} display the evolution and size distribution 
% the dynamics, evolution, and properties 
of all plasmoids considered in our calculations. In Fig.~\ref{fig:tracks}, we plot the temporal evolution of the plasmoids' centers for the three magnetizations as they evolve within the reconnection layer while the colour coding for each track denotes the plasmoid size $w_{\perp}/L$. Note that the duration of the adopted PIC simulations is different for the three values of $\sigma$. This will directly affect the observed duration of the reconnection event (see Sec.~\ref{sec:results}). Fig.~\ref{fig:final_size_hist} displays a histogram of the plasmoids' final\footnote{By \textit{final}, we refer to the value of any plasmoid property attained at the end of a plasmoid's lifetime. This could either be escape from the reconnection layer or merger with a neighbouring plasmoid.} transverse size, normalized to $L$. 
% just as they either advect from the reconnection layer or merge with another plasmoid. T
The dashed lines denote all plasmoids formed in a PIC simulation while the solid lines correspond to those plasmoids which pass our selection criterion. It can be seen in these two figures that the majority of the selected plasmoids, for all $\sigma$, have a final transverse size of $w_{\perp, f} \sim 0.01 L$. As will be discussed in Sec.~\ref{sec:results}, these plasmoids play an important role in the cumulative emission from a single reconnection event. Also present within the reconnection layer is a small number of \textit{monster} plasmoids with transverse sizes reaching $w_{\perp, f} \sim 0.1 \, L$ \citep{uzdensky2010}. These few plasmoids are crucial in shaping the overall light curve of a single reconnection event. 

\section{Model Description}
\label{sec:model} 
Let us consider a relativistic magnetized blazar jet with bulk Lorentz factor $\Gamma_{\rm j}$ and a half-opening angle $\theta_{\rm j} \sim 1/\Gamma_{\rm j}$. We assume that a fraction of the jet's energy is dissipated at a distance $z_{\rm diss}$ from the SMBH and radiated away \citep[but see also,][for continuous dissipation]{giannios2018}. The absolute power of a two-sided blazar jet can be estimated as \citep{celotti2007,dermer2009}:
\eqb
\label{eqn:L_j}
L_{\rm j} \approx 4 \pi (z_{\rm diss} \theta_{\rm j})^{2} c \beta_{\rm j} \Gamma_{\rm j}^{2} U_{\rm j}^{\prime},
\eqe
where $\beta_{\rm j} \approx 1$ is the dimensionless jet velocity, $z_{\rm diss}\theta_{\rm j}$ is the jet's cross-sectional radius, and $U_{\rm j}^{\prime}$ is the jet's energy density as measured in its co-moving frame. The latter is approximately equal to half of the magnetic energy density of the jet $U_{\rm j}^{\prime} \sim B_{\rm up}^{2} / 8\pi$, where $B_{\rm up}$ is the magnetic field strength of the regions far upstream from the reconnection layer, shown in SGP16 to be $\sim \sqrt{2}$ less than the area-averaged field strength within the plasmoids, $B$.  %\MP{Shouldn't be $2\pi$ in equation (1)? Are you assuming something for the pressure and energy density?} %\MP{Where does the 4 come in equation (1)? moreover, if $B$ is the unreconnected magnetic field, why not $B^2/8\pi$?} 

Energy dissipation in magnetized jets may occur via magnetic reconnection, which, in turn, can be triggered through various mechanisms; e.g., magnetic kink-instabilities \citep{spruit2001,giannios2006,barniol2017} or inversions in the vertical magnetic flux within the accretion disk \citep{parfrey2015,giannios2018}. Regardless of the formation process, we assume that the reconnection layer is a slab, of length\footnote{The length of the reconnection region is taken to be $2 \, L$ in the direction of the plasmoid's motion and $\sim L$ in the other two directions.} $2\, L$. The location of the dissipation region is widely debated, ranging from sub-pc scales \citep{ghisellini2009} to multi-pc scales \citep{tavecchio2010,costamante2018}. Henceforth, we adopt $z_{\rm diss} = 5\times10^{17}$~cm and $L=5\times10^{16}$~cm as indicative values.  

A schematic of our model is presented in Fig.~\ref{fig:layer_in_jet_sketch}. All results will be mapped to the frame of an observer, positioned at an angle $\theta_{\rm obs}$ with respect to the jet axis. Here, we consider values of $\theta_{\rm obs}$ from $0^{\circ}$ to $20^{\circ}$. Because the orientation of the reconnection layer in the jet is not known {\sl a priori}, we also vary the angle $\theta^{\prime}$ from $0^{\circ} - 180^{\circ}$. 

\subsection{Radiative Transfer Model}
\label{sec:radiation_transfer}
To track the temporal evolution of the particle distribution within each plasmoid and compute its photon emission, we numerically solve the following integro-differential equations for each plasmoid:
\eqb
\label{eqn:particle_kinetic}
\partial_{t} N^{e} = Q_{\rm inj}^{e} + Q_{\gamma\gamma}^{e} - L_{\rm syn}^{e} - L_{\rm IC}, \\
\label{eqn:photon_kinetic}
\partial_{t} N^{\gamma} + \frac{N^{\gamma}}{t_{\rm esc}} = Q_{\rm syn}^{\gamma} + Q_{\rm IC}^{\gamma} - L_{\gamma \gamma}^{\gamma} - L_{\rm ssa}^{\gamma}.
\eqe
Here, $N^{e}$ ($N^{\gamma}$, respectively) denotes the number of relativistic pairs (photons) having energies between $\gamma \rightarrow \gamma + {\rm d}\gamma$ ($x \rightarrow x + {\rm d}x$, where $x \equiv h \nu /m_{\rm e} c^{2}$) at time $t$, as measured in the plasmoid's co-moving frame. Equations~\ref{eqn:particle_kinetic} and~\ref{eqn:photon_kinetic} are solved numerically in the plasmoid's co-moving frame by following a similar method provided by \cite{chiaberge1999} (we refer the reader there for a more detailed description of the numerical scheme). There is no energy-gain term in eqn.~\ref{eqn:particle_kinetic}, since we assume the radiating particles have been accelerated before entering the plasmoid. Particle acceleration can take place at the X-points (i.e. the local minimum of the magnetic vector potential) by the reconnection electric fields or during mergers or plasmoid compression \citep{zhang2018,petropoulou2018} (see also Sect.~\ref{sec:discussion}). %\MP{In our paper with Lorenzo we talk about a different  process (compression)}. 
The operators $L$ and $Q$ denote the loss and source terms for pairs and photons. The radiative processes we consider are:

\begin{enumerate}

\item \textit{Synchrotron radiation}: The electron loss term is provided by eqn.~34 in \cite{mastichiadis1995} while the photon source term is derived using the full emissivity expression \citep{rybicki1986}.

\item \textit{Synchrotron Self Absorption}: Photon absorption is expected to dominate at low frequencies and the corresponding loss term is given as $L_{\rm ssa}^{\gamma} = w_{\perp} \alpha_{\nu} N^\gamma(x,t)/2$, where the absorption coefficient $\alpha_{\nu}$ is determined using the expressions in \cite{rybicki1986}. 

\item \textit{Inverse Compton Scattering}: We consider scatterings occurring in both the Thomson and Klein-Nishina regimes \citep[see eqns.~42--46 in][]{mastichiadis1995}. The seed photons for inverse Compton scattering can be internal (e.g. synchrotron photons produced by the relativistic pairs) and/or external (e.g. photons from the BLR) to each plasmoid. Here, we consider the BLR as the sole source of external photons, which is taken to be a blackbody emitter with peak frequency $\sim 5$~eV. The energy density of external photons is computed in the plasmoid's rest frame using the appropriate $\Gamma_{\rm p}$ for the Lorentz transformation, where $\Gamma_{\rm p}$ is the plasmoid's Lorentz factor as measured in the rest frame of the host galaxy (see Sec.~\ref{sec:results}). \citep[see also][]{ghisellini1996}. 
% The expressions used for the spectral energy density of the external photons are provided in Section~2.5 of \cite{ghisellini1996} while 
The values of the bolometric luminosity of the BLR $L_{\rm BLR}$ are provided in Table~\ref{table:model_parameters_2}. Other external radiation fields (e.g. from the accretion disk or the dusty torus), which may become relevant depending on the location of the reconnection layer within the jet, can also be included in the radiative transfer code. Our model allows for the addition of any external radiation source and is not limited to a blackbody emitter.

\item \textit{Photon-Photon Pair Production}: The loss and source terms for this process are given by eqns.~54 and 57 in \cite{mastichiadis1995}. Contributions from external photons emitted by the BLR are also included in this process. %\MP{Where does the last conclusion come from? did you compute the optical depth for attenuation of the escaping gamma-ray spectrum from the plasmoid as it propagates through the BLR?} \IC{It came from numerical estimates we performed months ago. Yes I did and it is negligible.}
\begin{table*}
\centering
\caption{Parameters used within our radiative transfer model. From left to right: magnetization $\sigma$, the blazar subclass, model name, slope $p$ of the injected particle distribution, the pair multiplicity $N_{\pm}$, minimum and maximum Lorentz factor of the injected particle distribution averaged over all plasmoids, magnetic field strength $B$ within all plasmoids, bolometric luminosity of the BLR $L_{\rm BLR, 45}$ normalized to $10^{45}$~erg~s$^{-1}$, bulk Lorentz factor of the blazar jet $\Gamma_{\rm j}$, the value of the ratio of the energy density of the injected pairs to the magnetic energy density $U_{\rm inj}^{\rm e}/U_{\rm B}$ averaged over all plasmoids and assuming no radiative losses, and the jet luminosity $L_{\rm j, 46}$ normalized to $10^{46}$~erg~s$^{-1}$. All parameters listed here are free except for $L_{\rm j}$ and $U_{\rm inj}^{e}$ which are determined from eqns.~\ref{eqn:L_j} and~\ref{eqn:U_e}, respectively. In all cases, the half-length of the layer is fixed at $L = 5 \times 10^{16}$~cm.}
\begin{threeparttable}
\begin{tabular}{ cccccccccccc }
\hline
$\sigma$ & Blazar & Model & $p$\tnote{a} & $ N_{\pm}$\tnote{b} & $\gamma_{\rm min}$\tnote{c} & $\gamma_{\rm max}$\tnote{d} & $B$~(G) & $L_{\rm BLR, 45}$~(erg~s$^{-1}$) & $\Gamma_{\rm j}$ & $U_{\rm inj}^{\rm e}/U_{\rm B}$ & $L_{\rm j, 46} \, ($erg~s$^{-1})$ \\
 & Class & Name & & & & & & & & &  \\
\hline
$3$ & FSRQ & F3 & $3$ & $7$ & $108$ & $5\times 10^3$ & $7$ & $12$ & $15$ & $2.8$ & $9$ \\
\hline
$10$ & FSRQ & F10 & $2.1$ & $6$ & $94$ & $5\times10^3$ & $7$ & $12$ & $12$ & $1.1$ & $9$\\
\hline
$10$ & BL Lac & B10 & $2.1$ & $1$ & $560$ & $5\times10^{4}$ & $2$ & $5\times 10^{-4}$ & $12$ & $1.4$ & $0.8$\\
\hline
$50$ & BL Lac & B50 & $1.5$ & $100$ & $1.1$ & $2.8\times10^4$ & $4$ & $5\times 10^{-4}$ & $10$ & $0.2$ & $3$\\
\hline
\end{tabular}
\tnote{a} Determined from PIC simulations of pair plasmas \citep{sironi2014,guo2014,werner2016}.\\
\tnote{b} Used in estimating the characteristic/average Lorentz factor of the injected particle distribution (see Appendix~\ref{sec:particle_distribution}).\\
\tnote{c} Determined using eqn.~\ref{eqn:g_min} for $\sigma = 3$ and $10$.\\
\tnote{d} Determined using eqn.~\ref{eqn:g_max} for $\sigma = 50$.
\end{threeparttable}
\label{table:model_parameters_2}
\end{table*}

\item \textit{Photon Escape}: When solving eqn.~\ref{eqn:photon_kinetic} for $N^{\gamma}$, we assume a fraction of the photons within a given plasmoid are escaping on a characteristic timescale of $t_{\rm esc} = w_{\perp}/2 c$. Light travel time effects due to the motion of plasmoids in the layer with respect to the observer are taken into account in our definitions of the plasmoid Doppler factor and the corresponding observer time (see Sec.~\ref{sec:results} and Appendix~\ref{sec:observer_time}, respectively). However, we do not take into account light travel time effects of photons within plasmoids.
\end{enumerate}

When solving eqn.~\ref{eqn:particle_kinetic} for the temporal evolution of the particle distribution within each plasmoid, we require the instantaneous injection rate of particles, $Q_{\rm inj}^{e} (\gamma, t)$. Along with the many properties of an individual plasmoid, $Q_{\rm inj}^{e} (\gamma, t)$ is also derived using PIC results. To determine this quantity, we require the instantaneous number of particles within a plasmoid at any given time to be $N_{\rm inst}(t) = V(t) \, n_{\rm co}$, where $n_{\rm co}$ is the co-moving particle number density within a given plasmoid. The number density is, in good approximation, constant in time (see panels~d-f in  Fig.~5 of SGP16). Its physical value is estimated by using the definition of $\sigma$:
\eqb
\label{eqn:physical_number_density}
n_{\rm co} \approx \frac{B^{2} \, n_{\rm PIC}}{16 \pi \sigma m_{\rm p} c^{2}}.
\eqe
Here, $n_{\rm PIC}/4$ is the ratio of the time-averaged (i.e. averaged over a plasmoid's lifetime) co-moving particle number density, obtained from PIC results, over the particle number density far upstream from the layer, given as $4$ particles per cell. %$B_{\rm up}$ is the magnetic field strength of the regions far upstream from the reconnection layer, shown in SGP16 to be $\sim \sqrt{2}$ less than the area averaged field strength within the plasmoids $B$. %\LS{need to explain a bit more. dimensionless with respect to what (I am guessing you mean the ratio between plasmoid and upstream density), time-averaged over what? also, can you remind the reader where the factor of 16 is coming from?}

Assuming the injected particle distribution is in the form of a power-law with slope $p$ (see \cite{sironi2014,guo2014,werner2016} for their model fitting and Table~\ref{table:model_parameters_2} for our adopted values) between $\gamma_{\rm min}$ and $\gamma_{\rm max}$, the instantaneous injection rate is written as:
\eqb
\label{eqn:q_inj}
Q_{\rm inj}(\gamma, t) = \frac{(1-p) \, \gamma^{-p}}{\gamma_{\rm max}^{1-p} - \gamma_{\rm min}^{1-p}} \, \partial_t N_{\rm inst}(t),
\eqe
where the differentiation is taken with respect to the time as measured in the co-moving frame of a plasmoid. The values of $\gamma_{\rm min}$ and $\gamma_{\rm max}$ are either prescribed manually or determined from PIC simulations, depending upon the slope of the injected particle distribution (see Appendix~\ref{sec:particle_distribution} for more details).
Rough equipartition between relativistic particles and magnetic fields is a direct result found in PIC simulations of relativistic reconnection \citep{spg15}. The energy density of the injected particles can be determined by using eqns.~\ref{eqn:physical_number_density} and~\ref{eqn:q_inj} and is given as (normalized to the magnetic energy density within the plasmoids):
\eqb
\label{eqn:U_e}
\frac{U_{\rm inj}^{e}}{U_{\rm B}} = \frac{1-p}{2-p} \, \frac{\gamma_{\rm max}^{2-p} - \gamma_{\rm min}^{2-p}}{\gamma_{\rm max}^{1-p} - \gamma_{\rm min}^{1-p}} \, \frac{n_{\rm PIC} m_{\rm e}}{2 \sigma m_{\rm p}}.
\eqe
%\MP{In he definition of jet power we use different symbols for the energy densities. Let's use same notation}. 
In Table~\ref{table:model_parameters_2}, we provide this ratio averaged over all plasmoids and determined without radiative cooling.
Eqns.~\ref{eqn:particle_kinetic} and~\ref{eqn:photon_kinetic} are continuously solved while the plasmoid is within the reconnection layer, using the appropriate particle injection rate described above. Following the merger of a plasmoid with its neighbor or advection from the layer, we continue to solve the integro-differential equations, for a fraction of a dynamical time $L/c$, while ceasing particle injection and keeping its size and magnetic field strength fixed.

\subsection{Model parameters}
One of our goals is to model the main blazar subclasses, namely flat spectrum radio quasars (FSRQs) and BL Lac objects. These two subclasses are traditionally differentiated by the spectral lines measured in their optical component \citep{urry1995}, but also have distinct features observed in their broadband spectral energy distributions (SEDs) (see also \cite{padovani2017} for a recent review). The SEDs of both subclasses have a double-hump feature, with the low-energy component associated with synchrotron radiation while the high-energy component is thought to originate from the inverse Compton scattering of either synchrotron photons (i.e. synchrotron self-Compton or SSC) or photons produced externally from the jet (i.e. external Compton or EC). BL Lac objects are observed to have broader spectra in both energy bands, while FSRQs are measured to have steeper spectral slopes in their low-energy component (see \cite{ackermann2015} for properties of \textit{Fermi}-detected blazars).

{To account for the wide spectral properties of both classes, we vary some free parameters of our model, which are presented in Table~\ref{table:model_parameters_2} and summarized below: 
\begin{itemize}
\item The slope of the injected particle distribution $p$, although determined from PIC simulations, is a key model parameter in our differentiation of blazar subclasses. For low $\sigma$, the particle spectra are soft, i.e. $p > 2$, which may be appropriate for many FSRQs. For large $\sigma$, PIC simulations show that $p < 2$, which may be more relevant for modeling BL Lacs. For $\sigma = 10$, $p\sim 2.1$ and because it is somewhat of a transition between the two, we can model both subclasses.

\item The pair multiplicity $N_\pm$, i.e. the number of pairs per proton, is an unconstrained property of the plasma within blazar jets. It has been inferred from observations that blazar jets may be baryon loaded, containing several pairs per proton \citep{celotti2007,ghisellini2009,madejski2016}. In our work, $N_\pm$ is used in determining the energy density of injected pairs $U_{\rm inj}^{\rm e}$ within a plasmoid as well as the minimum and maximum Lorentz factors of the injected particle distribution, $\gamma_{\rm min}$ and $\gamma_{\rm max}$. By varying $N_\pm$, we are changing the energy range of the injected particle distribution such that the resulting spectra are similar to those of the two main  blazar subclasses. For those plasma magnetizations where $\gamma_{\rm min}$ or $\gamma_{\rm max}$ are not directly determined by PIC simulations (see Appendix~\ref{sec:particle_distribution} and Table~\ref{table:model_parameters_2}), we fix their values so that the cutoff frequencies of the low or high-energy components of the SED are similar to the observed ones. 
% which are prescribed manually
% , their values are similarly chosen to obtain appropriate cutoff frequencies observed in the low and high-energy components of a blazar's SED.

\item For a fixed $\sigma$, the magnetic strength $B$ determines the total number of injected particles within each plasmoid (see eqn.~\ref{eqn:physical_number_density}) as well as several spectral features (e.g. cutoff frequency and peak luminosity in synchrotron spectra). We vary this parameter such that the flare luminosity is comparable to those measured in characteristic FSRQs and BL Lacs. 

\item The bolometric luminosity of the BLR, $L_{\rm BLR}$ implies a radius $R_{\rm BLR} \approx 10^{17} \,\xi^{-1/2}_{-1} L_{\rm BLR, 44}^{1/2}$~cm, where $\xi=0.1\xi_{-1}$ is the fraction of the disk luminosity that is being reprocessed by the BLR\footnote{Henceforth, $Q_{\rm X}$ denotes normalization to $10^{\rm X}$ in cgs units, unless stated otherwise.}. For small $L_{\rm BLR}$, the dissipation region falls outside the BLR (i.e. $z_{\rm diss} > R_{\rm BLR}$). This results in a de-boosting of the BLR energy density as measured in a plasmoid's rest frame. We adopt this case for BL Lac sources, as observations  suggest a very weak BLR, whose emission is being swamped by the jet's emission \citep{padovani2017}. For large values of $L_{\rm BLR}$, the dissipation region   is contained within the BLR, resulting in a boosting of the external photons in the plasmoid's co-moving frame. This is required in order to obtain a large EC component in the plasmoid's observed spectra. This is a key feature for the calculation of emission from FSRQs in our model.
\end{itemize}

Additionally in Table~\ref{table:model_parameters_2}, we provide the jet luminosity, determined from eqn.~\ref{eqn:L_j}, at the adopted dissipation distance $z_{\rm diss}=5\times10^{17}$~cm. %ratio of the energy density of the injected particles, averaged over all plasmoids and determined without radiative cooling, to the magnetic energy density. Rough equipartition between relativistic particles and magnetic fields is a direct result found in PIC simulations of relativistic reconnection \citep{spg15}. Also displayed is the jet luminosity, determined from eqn.~\ref{eqn:L_j}, at the adopted dissipation distance $z_{\rm diss}=5\times10^{17}$~cm. 

\section{Results}
\label{sec:results}
%%%%%% FIGURE BEGIN %%%%%%%%
\begin{figure}
\centering
\includegraphics[height=0.33\textwidth]{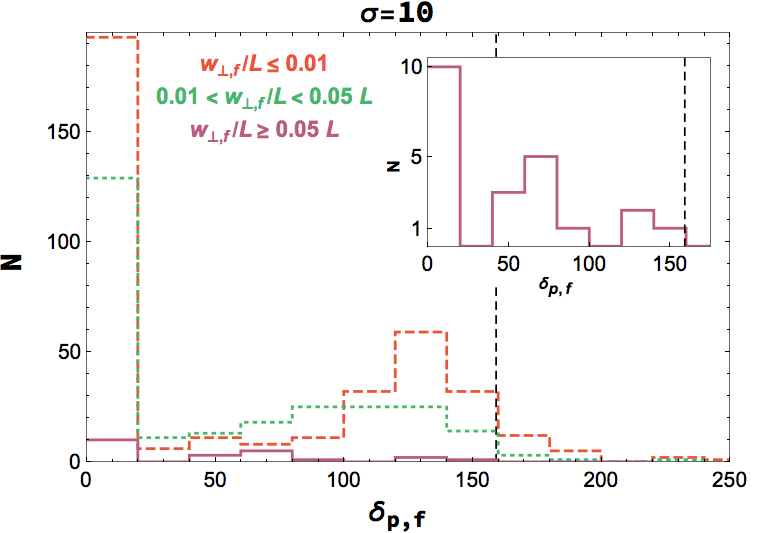}
\caption{Histograms of the final Doppler factors for all plasmoids for perfect alignment (i.e. $\theta_{\rm obs} = \theta^{\prime} = 0^{\circ}$) as produced from a $\sigma = 10$ PIC simulation in different bins of their final sizes: $w_{\perp, f}/L\le 0.01$ (dashed line), $0.01<w_{\perp, f}/L < 0.05$ (dotted line), and $w_{\perp,f}/L \geq 0.05$ (solid line); the latter is also displayed in the inset plot for clarity. 
% (see colour legend). The inset plot displays the histogram for plasmoids with final sizes $w_{\perp,f} \geq 0.05 L$. 
Due to the plasmoids' fast motion in the jet frame, their Doppler factor can be well in excess of the Doppler factor $\delta_{\rm j} \sim 2 \Gamma_{\rm j}$ expected from the jet motion alone. The dashed lines denote the value of the Doppler factor $\sim 4 \Gamma_{\rm j} \sqrt{1+\sigma}$, obtained from analytical estimates of \citet{giannios2009}, for fast plasmoids aligned with the observer. A coloured version of this plot is available online.}
\label{fig:cdf_final_doppler_factor}
\end{figure}
%%%%%% FIGURE END %%%%%%%%

By using the numerical setup described above and the parameters listed in Table~\ref{table:model_parameters_2}, we may numerically solve for the co-moving light curves and spectra for all plasmoids within the reconnection layer. Once determined, the results are then mapped to those seen by an observer by making the appropriate transformations using 
% $(\nu L_\nu)_{\rm obs} = \delta_{\rm p}^{4} (\nu L_\nu)_{\rm co}$, where 
the plasmoid's Doppler factor $\delta_{\rm p}$, which is written as:
\eqb
\label{eqn:Doppler_factor}
\delta_{\rm p} = \frac{1}{\Gamma_{\rm p} (1 - \beta_{\rm p} \cos (\theta-\theta_{\rm obs}))},
\eqe 
where $\Gamma_{\rm p}=\left(1-\beta_{\rm p}^2\right)^{-1/2}$ is the plasmoid's Lorentz factor as measured in the galaxy's frame. This is written as:
\eqb
\label{eqn:Gamma_p}
\Gamma_{\rm p} = \Gamma_{\rm j} \Gamma (1 + \beta_{\rm j} \beta \cos \theta^{\prime}),
\eqe 
where $\Gamma_{\rm j}$, $\beta_{\rm j}$ are the jet's bulk Lorentz factor and  dimensionless velocity, respectively, and $\beta$ is the plasmoid's velocity (in units of the speed of light) as measured in the reconnection frame. The angle  $\theta$ appearing in eqn.~(\ref{eqn:Doppler_factor}) is the angle between the plasmoid's direction of motion and the jet axis as measured in the galaxy's frame and is related to $\theta^\prime$ as:
% Here, $\theta$ is the angle of the plasmoid's dimensionless velocity, $\beta_{\rm p}$, with respect to the jet axis as measured by an observer. The angle $\theta$ is determined from
\eqb
\label{eqn:theta_1}
\tan \theta = \frac{\beta \, \sin \theta^{\prime}}{\Gamma_{\rm j} (\beta_{\rm j} + \beta \cos \theta^{\prime})}.
\eqe 

Some of our main results for $\delta_{\rm p}$ are displayed in Fig.~\ref{fig:cdf_final_doppler_factor}. Here, we plot histograms of the final Doppler factor for all plasmoids, as produced from a $\sigma = 10$ simulation in perfect alignment (i.e. $\theta_{\rm obs} = \theta^{\prime} = 0^{\circ}$), as a function of their final size. As seen in the plot and is applicable to all $\sigma$, we find that $\sim 1/2$ of all plasmoids have $\delta_{\rm p} \leq 2 \Gamma_{\rm j}$. This is due to plasmoids moving on the side of the reconnection layer which is oriented away from the observer (see green plasmoid in Fig.~\ref{fig:layer_in_jet_sketch}). Therefore, we expect for perfect alignment that only half of the total number of plasmoids will significantly contribute to the cumulative emission from the plasmoid chain. However, as will be shown in Sec.~\ref{sec:plasmoid_chain_results}, this is not always the case for other orientations. The remaining half of plasmoids in Fig.~\ref{fig:cdf_final_doppler_factor} with $\delta_{\rm p} \gtrsim 2 \Gamma_{\rm j}$ are spread over a large range of $\delta_{\rm p}$, with the spread increasing for larger $\sigma$. This is in part due to the increasing range of attainable $\Gamma$ for plasmoids in larger $\sigma$, up to the asymptotic value of $\sqrt{1+\sigma}$ (SGP16).

\subsection{Individual Plasmoids}
\label{sec:single_plasmoid_runs}
\begin{table}
\centering
\caption{Compton ratio for the three plasmoids of different $w_{\perp, f}$ presented in Figs.~\ref{fig:individual_spectra_plots_fsrq} and~\ref{fig:individual_spectra_plots_bl_lac}. The ratio of the peak components is determined at the time of the plasmoid flare's peak. }
\begin{tabular}{c|ccc}
\hline
Model & \multicolumn{3}{c}{$w_{\perp, f} / L$}\\
Name & $0.005$ &   $0.03$ &  $0.1$\\
\hline
F3 & $263$ & $162$ & $40.9$ \\
F10 & $222$ & $210$ & $28.8$ \\
B10 & $0.252$ & $0.106$ & $0.209$ \\
B50 & $0.0412$ & $0.111$ & $0.0719$ \\
\hline
\end{tabular}
\label{table:ec_to_syn_ratio}
\end{table}

%%%%%% FIGURE BEGIN %%%%%%%%
\begin{figure*}
\centering
\includegraphics[height=0.815\textwidth]{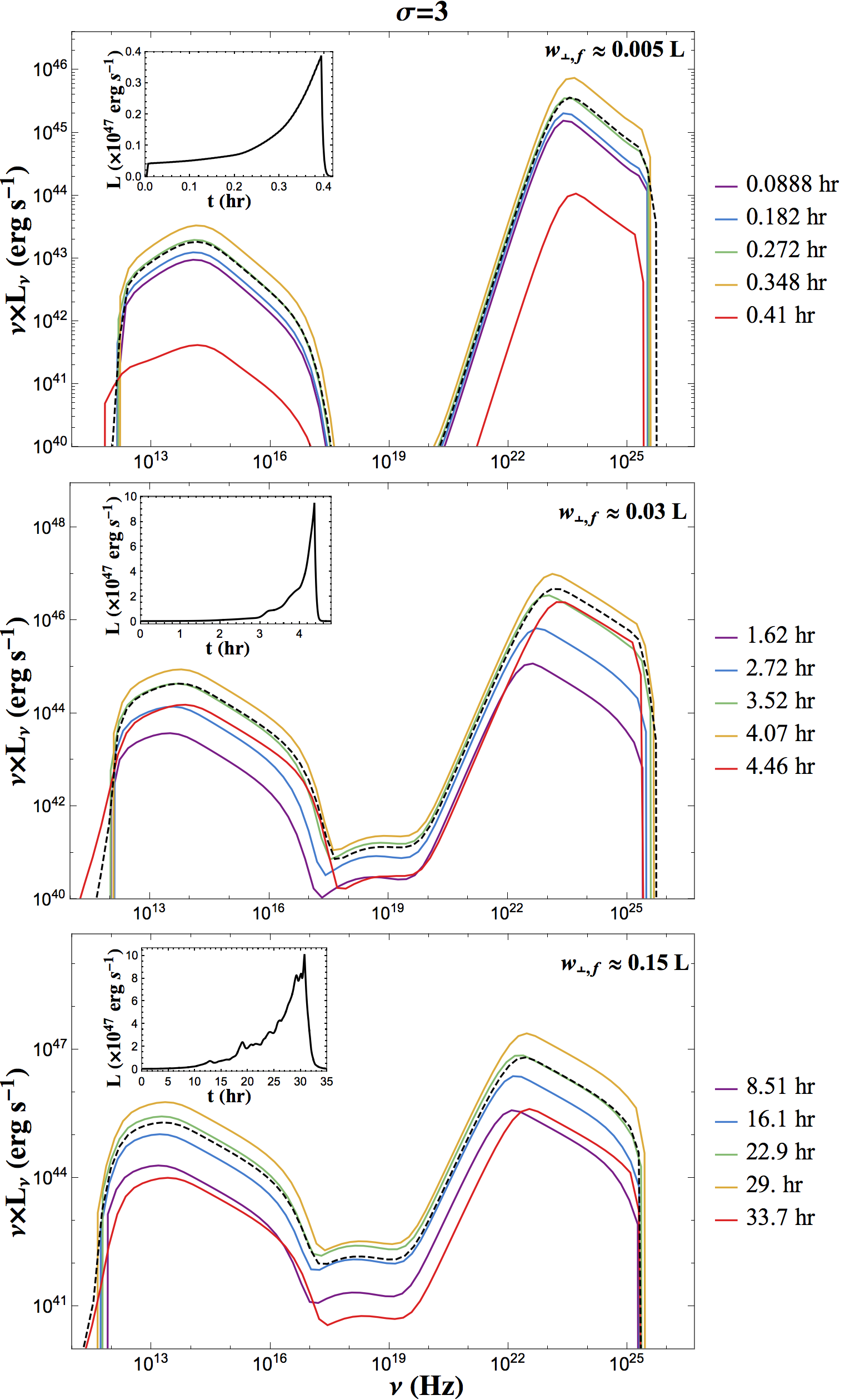}
\includegraphics[height=0.819\textwidth]{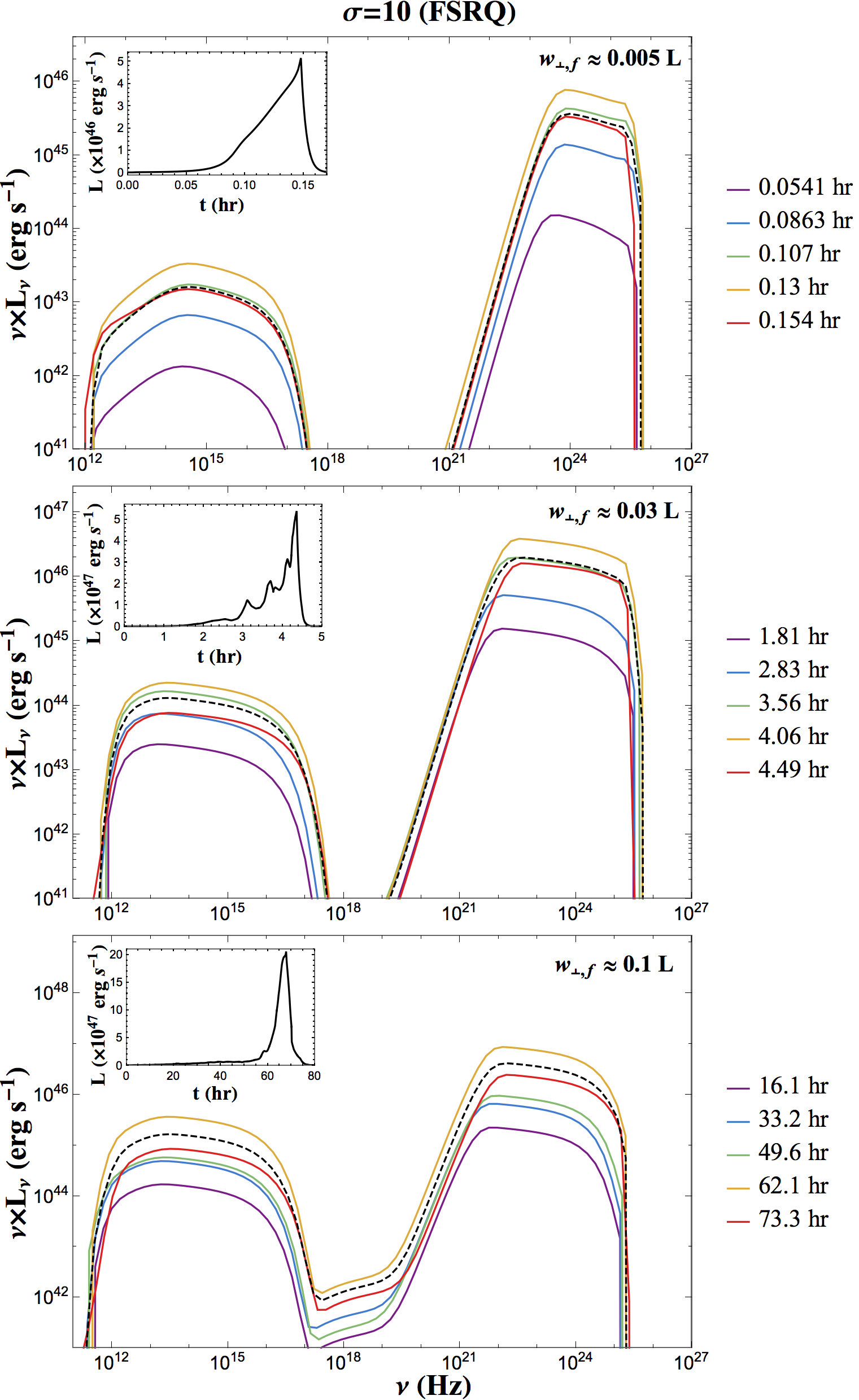}
\caption{Temporal evolution of plasmoids' broadband spectra for our modeling of an FSRQ-like source for two values of the jet's plasma magnetization: $\sigma=3$ (left column) and $\sigma=10$ (right column). For each column, the three panels correspond to the emission produced by a small, medium, and large plasmoid (from top to bottom), the final sizes of which are provided in each panel. The colour coding of the spectra indicates a time within the plasmoid's lifetime (see legend to right of each panel). The black dashed lines denote the time-averaged spectra of the respective plasmoid, averaged over the times listed in each legend. Within each panel is an inset plot of the bolometric light curve of the respective plasmoid presented in the spectra. All spectra and light curves are produced assuming perfect alignment (i.e. $\theta^\prime=\theta_{\rm obs}=0$). A coloured version of this plot is available online.}%\MP{I preferred the previous color scale you used.}}
\label{fig:individual_spectra_plots_fsrq}
\end{figure*}
%%%%%% FIGURE END %%%%%%%%

%%%%%% FIGURE BEGIN %%%%%%%%
\begin{figure*}
\centering
\includegraphics[height=0.815\textwidth]{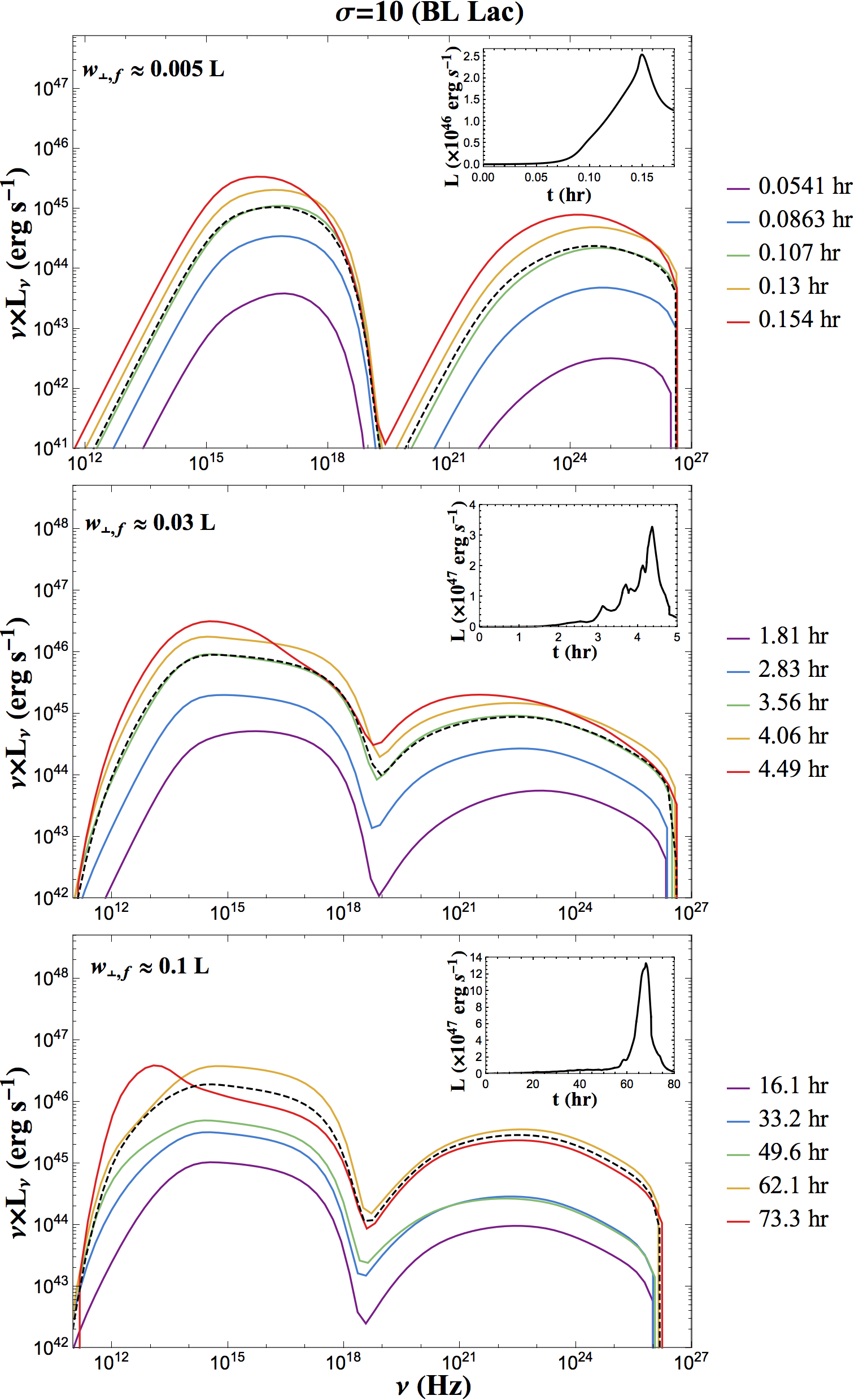}
\includegraphics[height=0.815\textwidth]{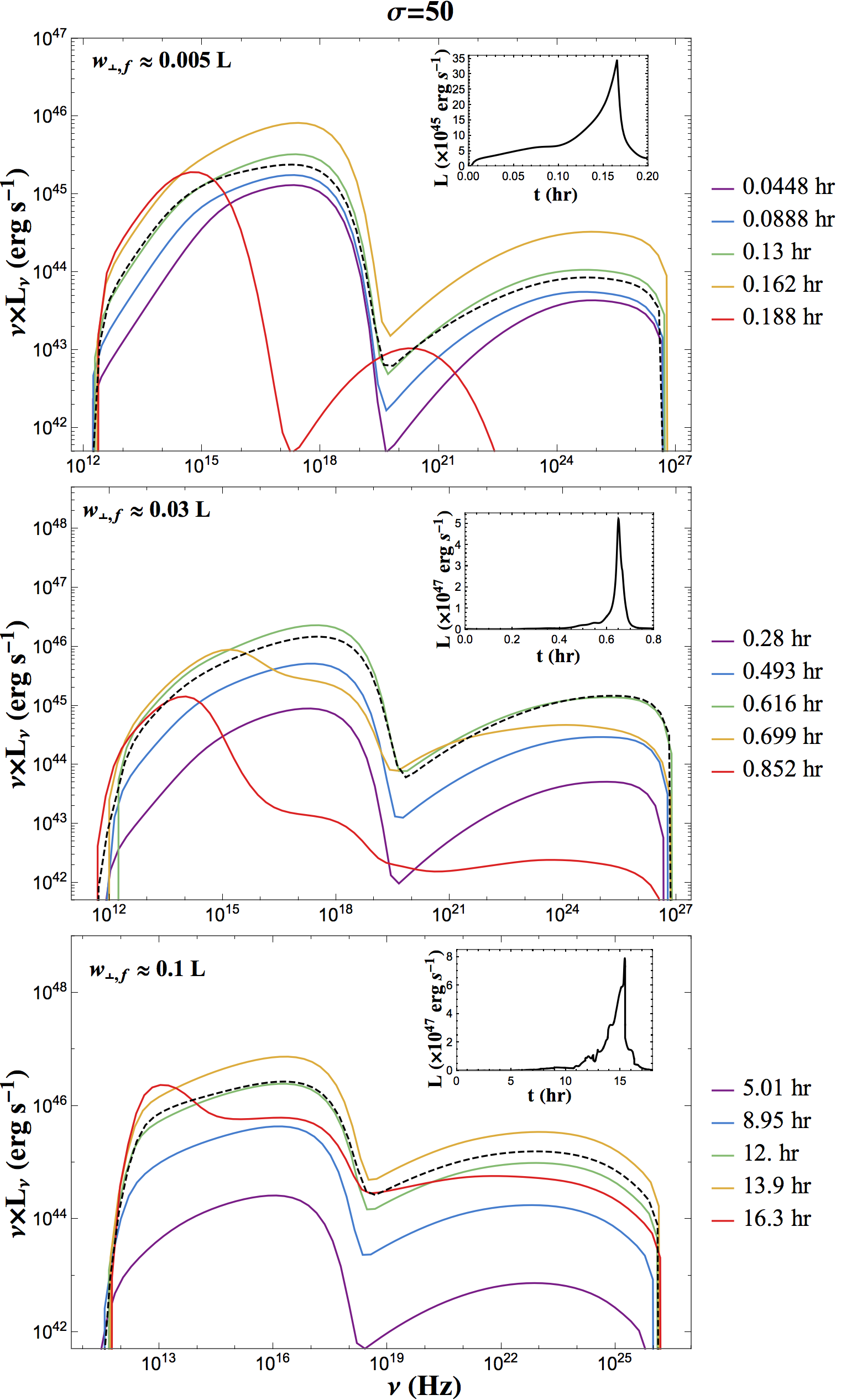}
\caption{Similar to Fig.~\ref{fig:individual_spectra_plots_fsrq} but for our modeling of a BL Lac-like source as produced from a $\sigma=10$ (left column) and $\sigma=50$ (right column) PIC simulation. All curves presented here are for perfect alignment. For some plasmoids, we find a low-energy bump in the SED occurring during the plasmoid's decay due to particles cooling to lower Lorentz factors (see bottom two panels at $\nu\sim10^{12}$~Hz). A coloured version of this plot is available online.}%\MP{I preferred the previous color scale you used.}}
\label{fig:individual_spectra_plots_bl_lac}
\end{figure*}
%%%%%% FIGURE END %%%%%%%%
As a first step, we compute the spectra and light curves for several individual plasmoids for each $\sigma$. Our results are displayed in Figs.~\ref{fig:individual_spectra_plots_fsrq} and~\ref{fig:individual_spectra_plots_bl_lac} for perfect alignment (i.e. $\theta_{\rm obs} = \theta^{\prime} = 0^{\circ}$). Fig.~\ref{fig:individual_spectra_plots_fsrq} displays our modeling of an FSRQ-like source, with the left and right columns denoting results from a $\sigma=3$ and $10$ PIC simulation, hereafter referred to as F3 and F10 respectively. Fig.~\ref{fig:individual_spectra_plots_bl_lac} shows our models of a BL Lac-like source, with the left and right columns corresponding to $\sigma = 10$ and $50$, which will hereafter be denoted as B10 and B50 respectively. For any column in the two figures, each panel corresponds to a plasmoid with a particular final transverse size; from top to bottom, we display a small ($w_{\perp, f} /L \sim 0.005$), a medium ($w_{\perp, f} /L \sim 0.03$), and a large ($w_{\perp, f} /L \sim 0.1$) plasmoid. The colour coding provided in the spectra plots denotes a particular time within the plasmoid's lifetime. The yellow curve denotes a time near a flare's peak, the red curve to a time following the peak of the flare, and the black dashed line denotes the time-averaged spectra of the respective plasmoid, averaged between the times provided in the legend of each spectra plot. Within each panel is an inset plot displaying the bolometric light curve for the plasmoid shown in the corresponding spectra plot.

A few things that are worth noting regarding the spectra and light curves and which hold regardless of the $\sigma$ value are listed below:
\begin{enumerate}

\item The general spectral shape of our results presented in Figs.~\ref{fig:individual_spectra_plots_fsrq} and~\ref{fig:individual_spectra_plots_bl_lac} have similar characteristics observed in FSRQs and BL Lacs. The former have steep spectral slopes (i.e. $p>2$) in the low-energy spectral component and a luminous high-energy EC component, while the latter has a very broad and flat spectrum. The spectral slope of the low-energy component of the SED motivated us to use the PIC results from low $\sigma$ to model FSRQs and results from high $\sigma$ to model BL Lac objects. In a majority of the presented spectra, we find a sharp cutoff occurring at both low and high frequencies due to the synchrotron self-absorption process and the Klein-Nishina cutoff, respectively. The location of these cutoffs is different for each plasmoid as the observed frequency is proportional to $\delta_{\rm p}$. As discussed in SGP16, small plasmoids are typically fast and large plasmoids are slow. This translates to higher cutoff frequencies in the photon spectra of smaller plasmoids and lower frequencies in those of larger plasmoids. 

\item For F3 and FSRQ 10, as shown in Fig.~\ref{fig:individual_spectra_plots_fsrq}, the peak luminosity ratio of the EC to synchrotron component decreases for increasing plasmoid size. The energy density of external photons from the BLR measured in the plasmoid's co-moving frame is $\Gamma_{\rm p}^{2} U_{\rm BLR}$, where $U_{\rm BLR}$ is measured in the rest frame of the SMBH. Because smaller plasmoids have larger Lorentz factors in the reconnection layer, the energy density of seed photons for Compton scattering (in their co-moving frame) is higher, and the power of the EC component is also larger. In Table~\ref{table:ec_to_syn_ratio}, we provide the ratio of the peak high-energy component (i.e. EC for F3 and F10 or SSC for B10 and B50) to the synchrotron peak, taken at the times a plasmoid flare has reached its peak luminosity. Still, the bolometric luminosity increases for increasing plasmoid size, due to the larger number of emitting particles (see also PGS16). For B10 and B50, one could increase the SSC to synchrotron ratio by decreasing the magnetic fields strength 
$B$. Interestingly, we obtain large values of the Compton dominance without requiring $U_e \gg U_b$, as inferred by one-zone models of FSRQ flares \citep{ackerman2016}.%\MP{Maybe mention that we get large values of the Compton dominance without requiring $U_e \gg U_b$, as inferred by one-zone models of FSRQ flares (e.g. give ref \url{http://adsabs.harvard.edu/abs/2016ApJ...824L..20A}}
%\MP{What one would need in order to make the ratio of order unity for BL Lacs? This is what is typically observed}

\item Small plasmoids, with final transverse size $w_{\perp, f} \lesssim 0.005 \, L$, produce very short duration flares with low luminosity as compared to the medium and large plasmoids. This is a direct result of their shorter lifetimes and lower number of injected particles, and provides a verification to our reasoning for neglecting the majority of these plasmoids (see Sec.~\ref{sec:plas_select}).

\item When comparing the light curves of the medium and large plasmoids, we find that the peak luminosities $L_{\rm pk}$ are comparable, but with the flare duration being smaller for the medium sized plasmoids. The latter fact is due to both the shorter lifetime of a medium sized plasmoid and the fact that these plasmoids move relativistically within the reconnection layer (see Fig.~10 in SGP16), resulting in large Doppler factors (see Fig.~\ref{fig:cdf_final_doppler_factor}) and short observation times. The largest plasmoids move almost non-relativistically in the layer (i.e., $\Gamma \sim 1$) and have long lifetimes (of a few dynamical times $L/c$), as measured in the reconnection frame, resulting in longer duration flares.

\item The rising segments of the light curves of the medium and large plasmoids are found to be variable as opposed to the small plasmoids, whose light curves are generally smooth increasing functions. The former plasmoids (i.e. medium and large sizes) undergo mergers with smaller neighbours which result in variations in their Lorentz factor $\Gamma$. These variations are then carried over to the Doppler factor $\delta_{\rm p}$ (see Fig.~\ref{fig:doppler_factor_t_obs} in Appendix~\ref{sec:observer_time}), resulting in structured light curves. 
% substructures in the light curve of an individual plasmoid. 

\item As can be seen in all light curves of Figs.~\ref{fig:individual_spectra_plots_fsrq} and~\ref{fig:individual_spectra_plots_bl_lac}, regardless of plasmoid size and blazar subclass, the decay is found to be rather steep. This is a direct result of abruptly terminating particle injection post merger or advection while keeping the plasmoid's magnetic field strength fixed. Our calculations, therefore, are not designed to realistically calculate the decline phase of the emission after the merger or ejection of plasmoids from the reconnection layer. For a calculation that includes adiabatic losses and decay of the magnetic field strength after the plasmoid ejection, see PGS16.  
\end{enumerate}

As shown in Figs.~\ref{fig:individual_spectra_plots_fsrq} and~\ref{fig:individual_spectra_plots_bl_lac}, an individual plasmoid can produce a multi-wavelength flare across the entire electromagnetic spectrum. A direct result of our model is that we expect flares to occur in the optical, UV, X-ray, and $\gamma$-ray bands. For our examples of BL Lac-like sources, we also find flares appearing at energies $\gtrsim 1$~TeV. This work is intended to display the general results of an entire reconnection event. A comparison of our model results to blazar observations will be provided elsewhere (Christie et al., in prep.).

\subsection{Plasmoid Chain}
\label{sec:plasmoid_chain_results}
Here, we continue by investigating the role of our model parameters in shaping the cumulative light curves and SEDs of an entire reconnection event. In doing so, we also compare our results with previous analytical estimates provided by PGS16 and \cite{pcsg17}.

%\MP{I would also suggest to have a short section after the light curves section, about the total SED from the layer at different times. Then, we can also provide link to the animated movies that we will upload. We can also overplot some indicative SEDs of observed flares. }
\subsubsection{Light curves}
\label{sec:lc}
%%%%%% FIGURE BEGIN %%%%%%%%
\begin{figure*}
\centering
\includegraphics[width=0.38\textwidth]{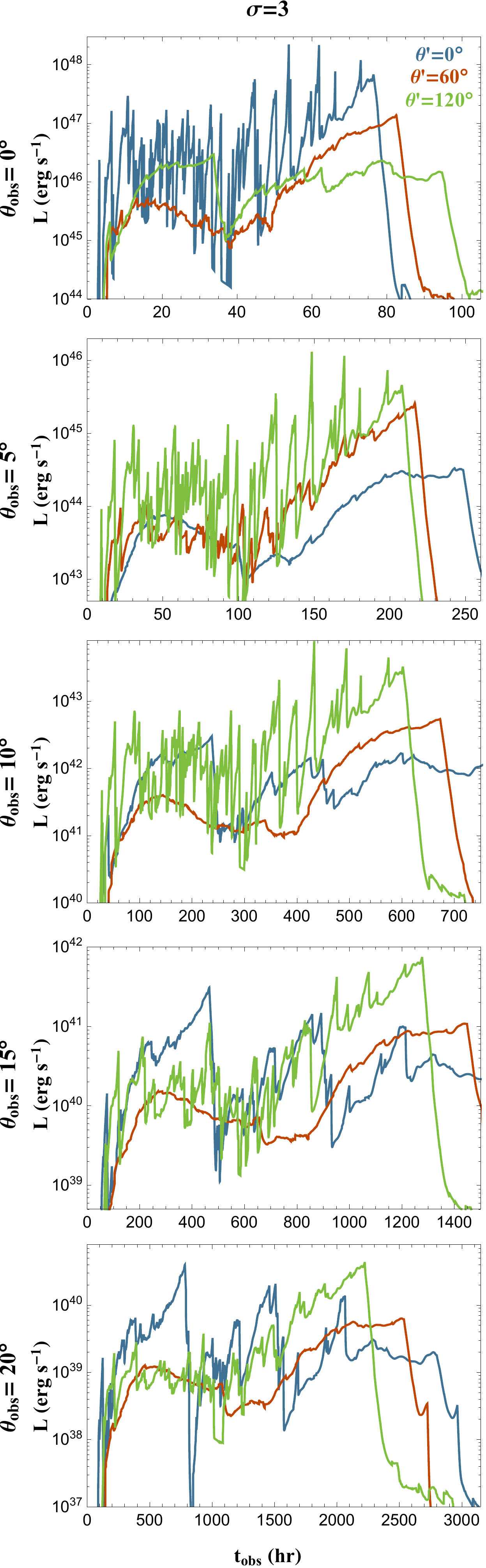}
\includegraphics[width=0.38\textwidth]{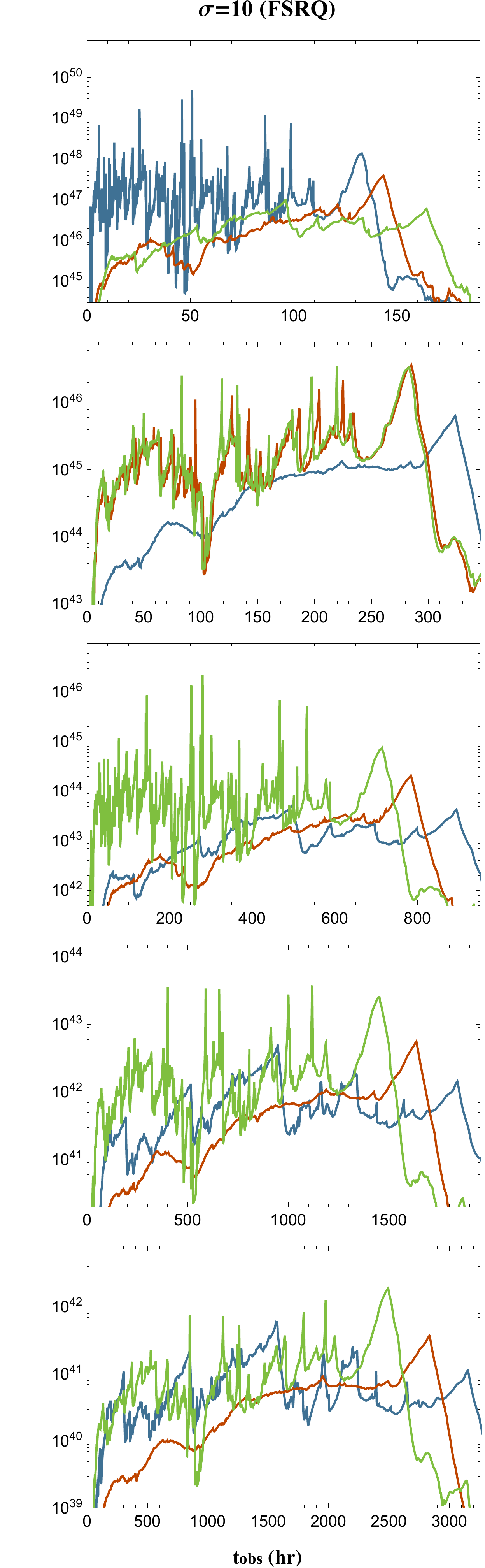}
\caption{Observed $\gamma$-ray ($0.1 - 300$~GeV) integrated light curves of an entire plasmoid chain for our modeling of a FSRQ as produced from a $\sigma=3$ (left panel) and a $\sigma=10$ (right panel) PIC simulation. Each row denotes a particular $\theta_{\rm obs}$ while the colour coding corresponds to different $\theta^{\prime}$ values, as denoted in the top left panel (a coloured version of this plot is available online). The high variability seen in several light curves, roughly one for each panel, results from maximizing the Doppler factor for all plasmoids. Those light curves which show little to no variability are produced solely from the largest plasmoids within the reconnection layer as the Doppler factor is $\delta_{\rm p} \sim \delta_{\rm j} \sim 1$ for all plasmoids, a direct result of the orientation. }%and the emission from a plasmoid is produced solely in its co-moving frame.}%Each column represents the results for a particular $\sigma$ value while each row denotes a different $\theta_{\rm obs}$ value. Within each panel, we plot four light curves, each colour denoting a different $\theta^\prime$ value (see top right panel for colour legend).}
\label{fig:gamma_ray_lc_fsrq}
\end{figure*}
%%%%%% FIGURE END %%%%%%%%

%%%%%% FIGURE BEGIN %%%%%%%%
\begin{figure*}
\centering
\includegraphics[width=0.3795\textwidth]{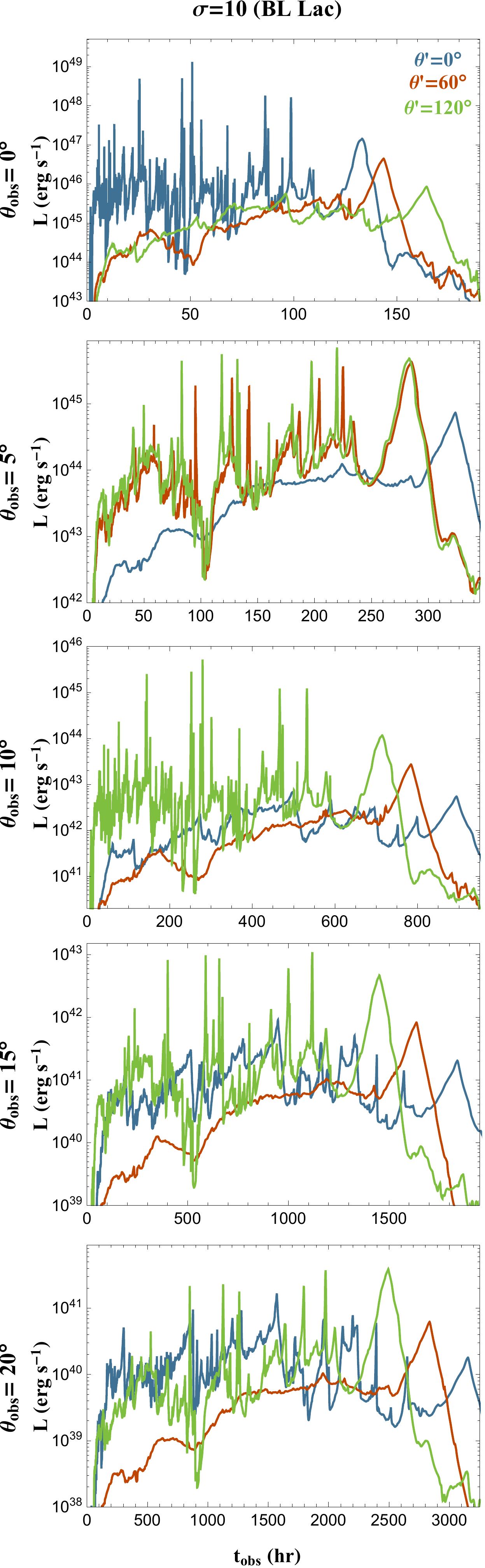}
\includegraphics[width=0.38\textwidth]{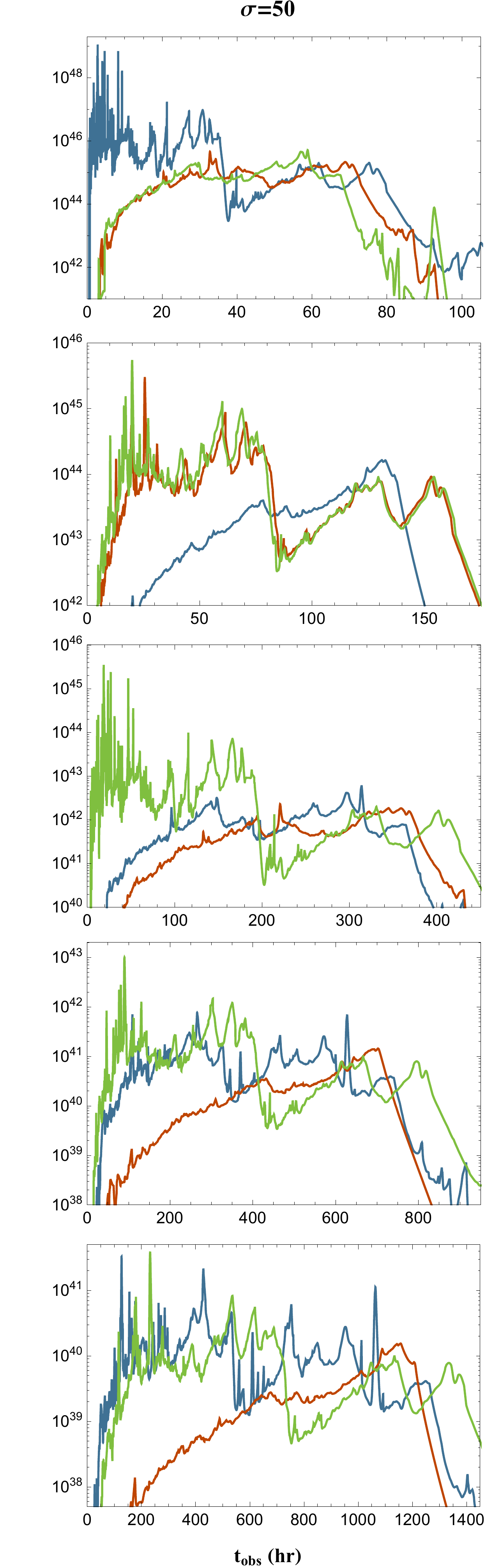}
\caption{Similar to Fig.~\ref{fig:gamma_ray_lc_fsrq} but for our modeling of a characteristic BL Lac using results from a $\sigma = 10$ (left panel) and a $\sigma=50$ (right panel) PIC simulation. A coloured version of this plot is available online.}
\label{fig:gamma_ray_lc_bllac}
\end{figure*}
%%%%%% FIGURE END %%%%%%%%

%%%%%% FIGURE BEGIN %%%%%%%%
\begin{figure*}
\centering
\includegraphics[width=0.38\textwidth]{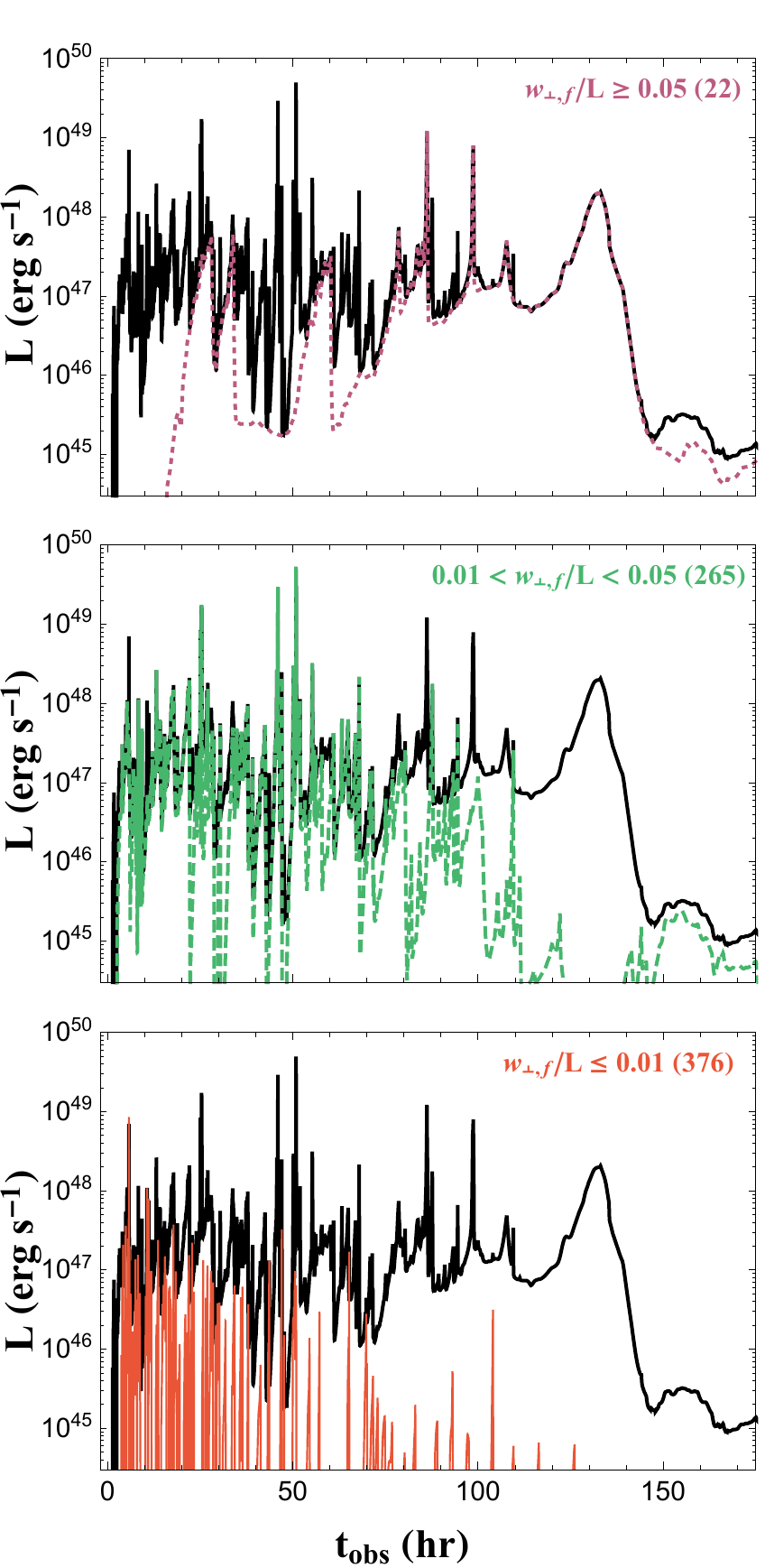} 
\hspace{0.75 cm}
\includegraphics[width=0.38\textwidth]{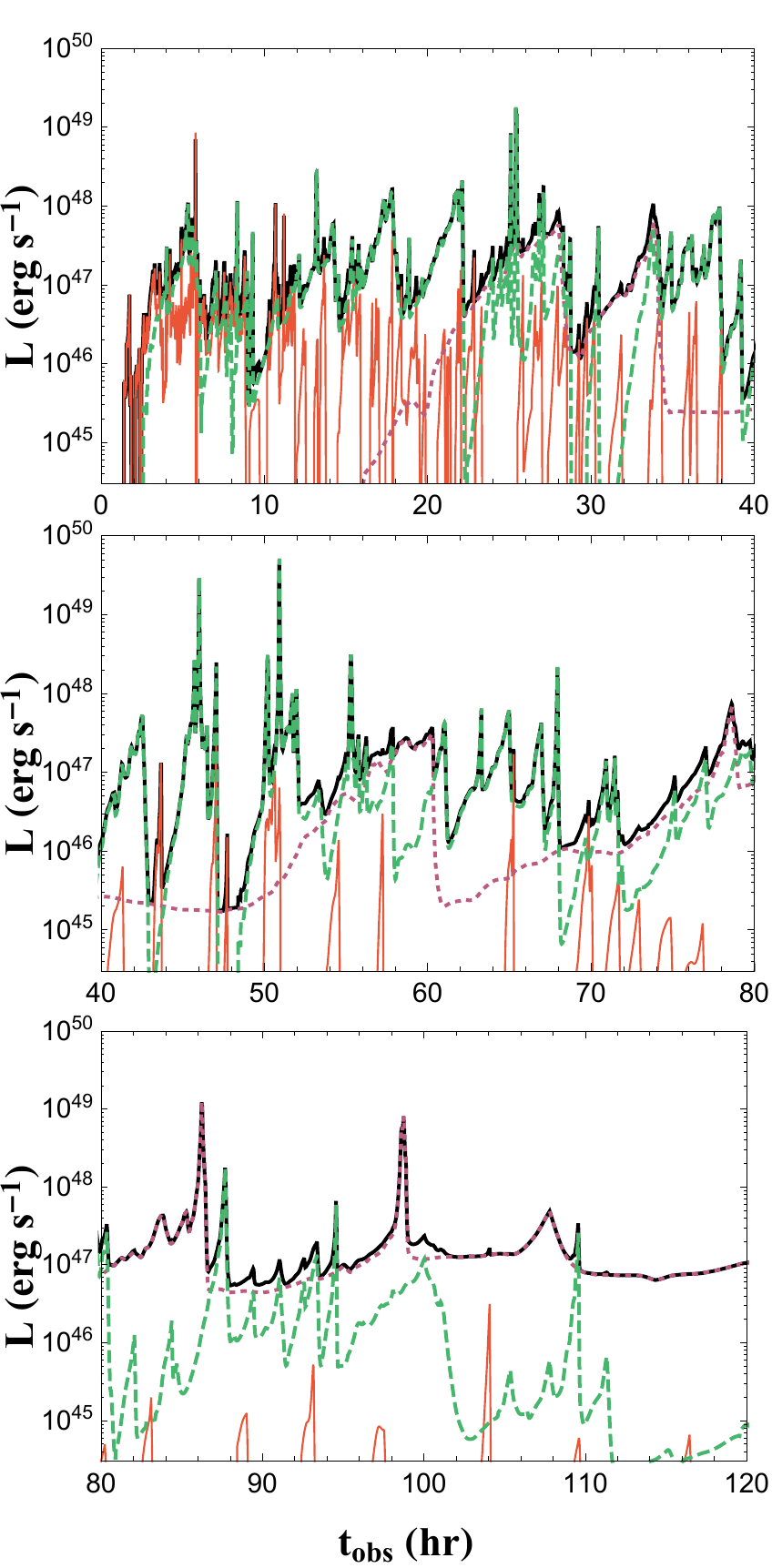}
\caption{Left Column: Breakdown of the bolometric light curve (solid black lines), for F10 with perfect alignment, as a function of the plasmoid's final transverse size $w_{\perp, f}$. From top to bottom, we present the contribution from plasmoids whose final sizes are within the ranges $w_{\perp,f}/L > 0.05$ (dotted lines), $0.01 < w_{\perp,f}/L < 0.05$ (dashed lines) and $w_{\perp,f}/L < 0.05$ (solid lines),
% (similar to Fig.~\ref{fig:cdf_final_doppler_factor}), 
respectively, while the total number of plasmoids within each range are provided in parenthesis. Right Column: Zoomed-in portions of the bolometric light curve, similar to the left column, separated in $40$~hr intervals. In both columns, the overall structure of the light curve is produced by the largest plasmoids while the rapid variability is due to medium sized plasmoids. A coloured version of this plot is available online.}
\label{fig:sigma_10_light_curve_breakdown}
\end{figure*}
%%%%%% FIGURE END %%%%%%%%

By combining the results of all individual plasmoids, presented above, we obtain a cumulative light curve for a single reconnection event. Our $\gamma$-ray integrated (i.e. $0.1 - 300$~GeV) light curves are presented in Figs.~\ref{fig:gamma_ray_lc_fsrq} for F3 and F10 and Fig.~\ref{fig:gamma_ray_lc_bllac} for B10 and B50. Each panel within the figures denotes a particular $\sigma$ and $\theta_{\rm obs}$ value while the colour scheme corresponds to a different $\theta^{\prime}$ (see legend in the top left panel of each figure). There are several features of our model which are key in understanding the shape, duration, and variability of the cumulative light curves and we discuss them below:

\begin{enumerate}
\item \textit{Orientation}: The orientation of the reconnection layer with respect to the jet axis and an observer's line of sight greatly affects the observed luminosity and duration of each individual plasmoid-powered flare. In the context of the entire plasmoid chain, this translates to the observed variability and overall duration of the reconnection process. For orientations resulting in 
% such that there is 
a large Doppler boosting in the emission of a small and medium sized plasmoids, the cumulative light curves appear highly variable and contain a superposition of both short and long duration flares. This feature can be seen in many of the cumulative light curves; e.g., for $\theta_{\rm obs} = \theta^{\prime} = 0^\circ$ or $\theta_{\rm obs} = 10^\circ$ and $\theta^{\prime} = 120^\circ$. For the other orientations, such that the Doppler factor for all plasmoids is $\delta_{\rm p} \sim \delta_{\rm j} \sim 1$ (e.g. $\theta_{\rm obs} = 10^\circ$ and $\theta^{\prime} = 60^\circ$), where $\delta_{\rm j}$ is the jet's Doppler factor, the short timescale variability is washed away and the overall shape of the cumulative light curves is determined by the largest plasmoids (i.e. final transverse size $w_{\perp, f} \sim 0.1 \, L$) within the layer, which contain most of the emitting particles. We stress that when the Doppler factors of a majority of the plasmoids is $\sim 1$, it is solely due to the particular orientation of the layer and observer with respect to the jet axis. The plasmoids themselves are characterized by fast motions within the layer, as shown in SGP16.

Besides variability, the orientation of the layer with respect to an observer affects the observed duration of the entire reconnection process. For increasing $\theta_{\rm obs}$, $\delta_{\rm p}$ decreases for all plasmoids, regardless of the $\theta^{\prime}$ value, thereby producing a longer duration reconnection event (see Appendix~\ref{sec:observer_time} for definition of the observer time). This can be seen for any $\sigma$ by moving from the top to bottom panels in either column of Figs.~\ref{fig:gamma_ray_lc_fsrq} or~\ref{fig:gamma_ray_lc_bllac}.\\

\item \textit{$\sigma$-value}: As reported by SGP16, the total number of secondary plasmoids produced within the reconnection layer increases for larger $\sigma$ (see Figs.~\ref{fig:tracks} and~\ref{fig:final_size_hist}). This allows for the production of more flares during the reconnection event.
% within the cumulative light curve. 
For example, this can clearly be seen by comparing the light curves of F3 and F10, shown in Fig.~\ref{fig:gamma_ray_lc_fsrq}. Because F10 contains more plasmoids than F3, we find a more variable light curve in the former. Additionally, the maximum $\Gamma$ attainable by plasmoids increases for larger $\sigma$. This in turn, results in higher $\delta_{\rm p}$, shorter observer times, and brighter flares for plasmoids.\\

\item \textit{Duration of PIC simulations}: Because we are directly using the results of SGP16, our radiative transfer calculations do not extend beyond the duration of each PIC simulation, which we identify as the duration of the reconnection event. The latter differs among the simulations with different $\sigma$, as shown in Fig.~\ref{fig:tracks}, and has an impact on the computed light curves.
It can be seen for any particular light curve presented in Figs.~\ref{fig:gamma_ray_lc_fsrq} and~\ref{fig:gamma_ray_lc_bllac} that there is a final flare at late times followed by a sharp decay. From this point onwards, an observer no longer receives emission from plasmoids in the  side of the reconnection layer oriented towards him/her, a result of the simulation's finite duration. Increasing the latter would result in the production of more plasmoids at this side of the layer and an extended reconnection event. Although the observer no longer sees events occurring on this side of the layer, he/she can receive radiation from plasmoids on the opposite side of the layer, as the radiation takes longer to reach the observer. This emission is under luminous and begins to appear after the major decay described above. The sharp decay, denoting the end of a reconnection event, can be seen at late times in any panel of Figs.~\ref{fig:gamma_ray_lc_fsrq} or~\ref{fig:gamma_ray_lc_bllac}; see e.g., green curve at $t \gtrsim 90$~hrs in the top left panel of Fig.~\ref{fig:gamma_ray_lc_fsrq} (i.e. F3, $\theta_{\rm obs} = 0^{\circ}$, and $\theta^{\prime} = 120^{\circ}$). 
% This is a common feature seen in all our presented light curves as we assume a fixed magnetic field strength within the plasmoids while ceasing particle injection. MP: Not necessary to mention again the reason. 

\item \textit{Plasmoid Size}: For select orientations where no rapid variability is seen in the cumulative light curves, the majority of the emission is produced from large plasmoids (i.e., $w_{\perp, f} \sim 0.1 \, L$), which are long-lived and their co-moving luminosities are orders of magnitude larger than those of the small and medium sized plasmoids. Because in these orientations $\delta_{\rm p} \sim 1$, the observed luminosities of individual plasmoid-powered flares are of the same order of magnitude as their co-moving luminosities. %\MP{Add some explanation about the non-relativistic plasmoids for which $\delta_p=\delta_j$. Then, you can say that this is of order unity for selected $\theta_{obs}$}

The variability observed in the cumulative light curves is related to the Doppler factor of plasmoids which, in turn, depends upon their size. As shown in Fig.~\ref{fig:cdf_final_doppler_factor}, the majority of medium sized plasmoids (i.e. $0.01 < w_{\perp, f} / L < 0.05$) from a $\sigma=10$ PIC simulation have $\delta_{\rm p} \sim 100$. These plasmoids therefore produce the short duration flares observed in many of the cumulative light curves. This is demonstrated in the middle panel of the left column of Fig.~\ref{fig:sigma_10_light_curve_breakdown}, where we plot a decomposition of the bolometric light curve into the contributions of plasmoids with different sizes.
% for our modeling of an FSRQ-like source using $\sigma=10$ PIC results and perfect alignment, as a function of the plasmoid's size.

The majority of the emission at early times is produced by small plasmoids with final sizes $w_{\perp, f} \leq 0.01 \, L$. Soon after the onset of the reconnection event,  
% When the PIC simulations of SGP16 are initialized and the tearing instabilities are developing, 
all plasmoids within the reconnection layer, regardless of $\sigma$, are born small. As the simulation develops, mergers begin to occur resulting in the formation of larger plasmoids. At later stages of the reconnection event, a \textit{monster} plasmoid (i.e. $w_{\perp, f} \sim 0.1 \, L$) can develop resulting in a long-duration event, shown as the last peak before the steep decay in the left column of Fig.~\ref{fig:sigma_10_light_curve_breakdown}. These select few plasmoids which can obtain large sizes provide the envelope for a large portion of the cumulative emission, as shown by all three panels in the right column of Fig.~\ref{fig:sigma_10_light_curve_breakdown}, where we display several zoomed-in portions of $40$~hr intervals of the bolometric light curve. Smaller sized plasmoids produce fast and brighter flares which appear on top of the larger plasmoid's flare/envelope. 

A common feature 
% As can be seen in a majority 
of the light curves presented in Figs.~\ref{fig:gamma_ray_lc_fsrq} and~\ref{fig:gamma_ray_lc_bllac} is that the reconnection event ends with a large plasmoid advecting from the layer (see Fig.~\ref{fig:tracks}). The expectation of a luminous, long-duration flare at the end of a blazar's flaring activity depends on whether the reconnection event last for long enough time to allow the formation of monster plasmoids. 
% The existence of a long duration flare at the end of each light curve is not a prediction of our model but rather a choice of when the PIC simulation was terminated. 
For example, if an observer was oriented on the left side of the layer (see Figs.~\ref{fig:tracks} and~\ref{fig:layer_in_jet_sketch}), for our $\sigma=3$ and $10$ results, he/she would observe numerous short duration flares following which a cutoff would occur at the end of the reconnection event.
\end{enumerate}

\subsubsection{Spectral Energy Distribution}
\label{sec:sed}
In addition to the light curves produced by a single reconnection event, we can also determine its SED. The spectral shape, its features, and temporal evolution parallel those of the individual plasmoids which constitute it (see Figs.~\ref{fig:individual_spectra_plots_fsrq}, \ref{fig:individual_spectra_plots_bl_lac} and Sec.~\ref{sec:single_plasmoid_runs} for a detailed description of an individual plasmoid's spectra). For this reason, we do not present the results here but provide it as supplementary material in the form of animated figures\footnote{\url{https://goo.gl/YDp2QM}}. In doing so, we provide the evolution of $\gamma$-ray integrated (i.e. $0.1-300$~GeV) light curves produce from F3, B10, and B50 in perfect alignment along with temporal snapshots of their respective SED. For F10, we provide a similar evolution for perfect and non-perfect (i.e. $\theta_{\rm obs} = 10^{\circ}$ and $\theta^{\prime} = 60^{\circ}$) alignment while including the temporal evolution of the individual plasmoids displayed in Fig.~\ref{fig:individual_spectra_plots_fsrq}.

A unique feature which appears in the temporal evolution of a cumulative SED is a high or low-energy component containing multiple peaks. The SED peak of plasmoids which are in the decay phase of their evolution is shifted to lower frequencies as particles contained within them cool to lower energies. A combination of this with the contributions from plasmoids in their growth phase provide this double-peak feature occurring in either the high or low-energy component. These features are a direct result of our simplified treatment of particle injection (or, lack of) and magnetic field evolution post-advection or merger. Regardless, it would be challenging to detect such fast evolving spectral features.
% with durations up to $\sim 1$~hr. \LS{not sure I understand why you need to quote 1 hr; based on what we say, this is actually not that small; i would remove  "with durations up to 1 hr"}}
% \MP{Let's comment also that these features are a direct result of our simplified treatment of the field evolution and particle injection after the end of the plasmoid's lifetime.}

\subsubsection{Decomposition of the light curves}
\label{sec:dist}
%%%%%% FIGURE BEGIN %%%%%%%%
\begin{figure*}
\centering
\includegraphics[width=0.75\textwidth]{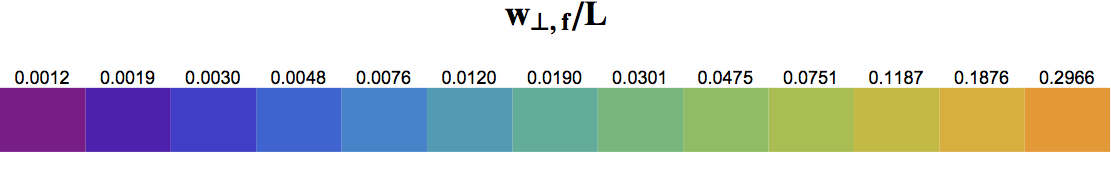}\\
    
\includegraphics[height=0.31\textwidth]{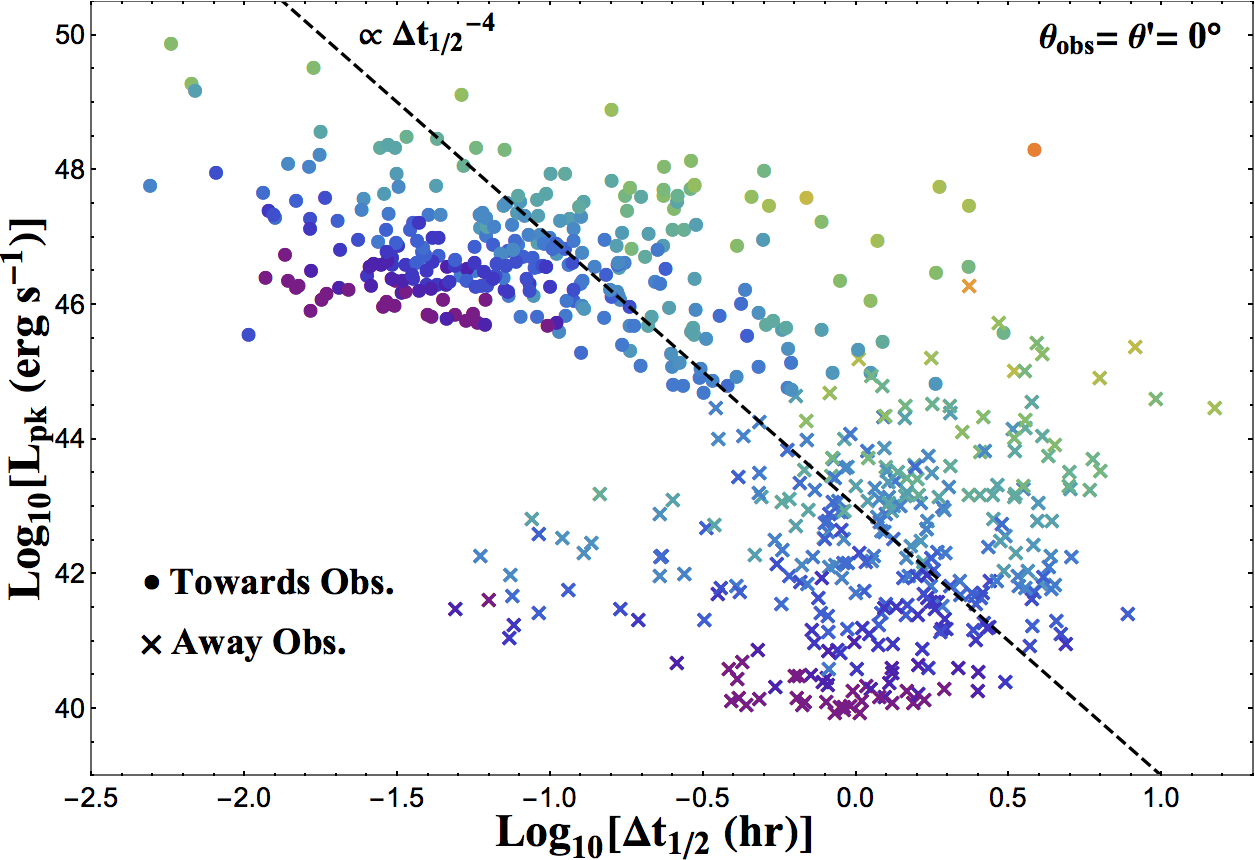}
\hspace{0.2cm}
\includegraphics[height=0.31\textwidth]{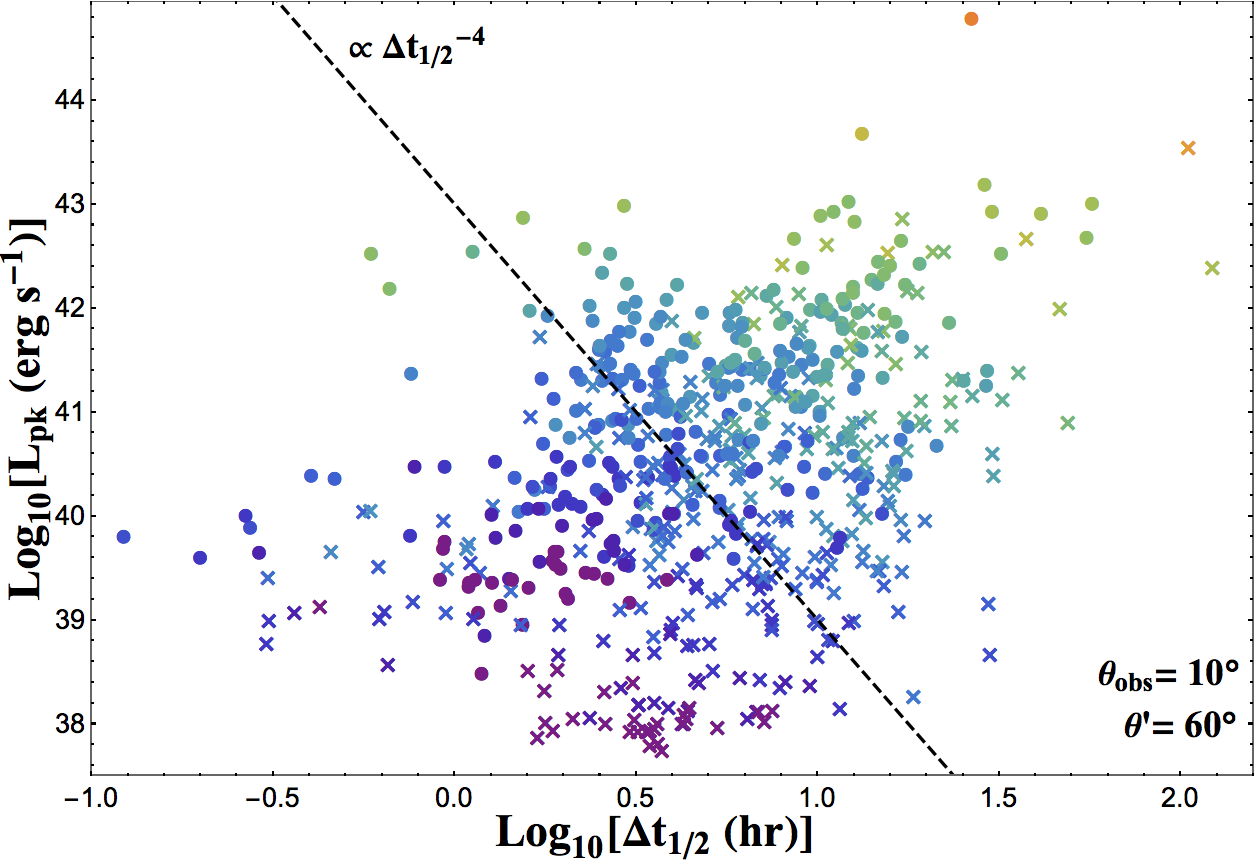}\\

\vspace{0.3cm}

\includegraphics[width=0.75\textwidth]{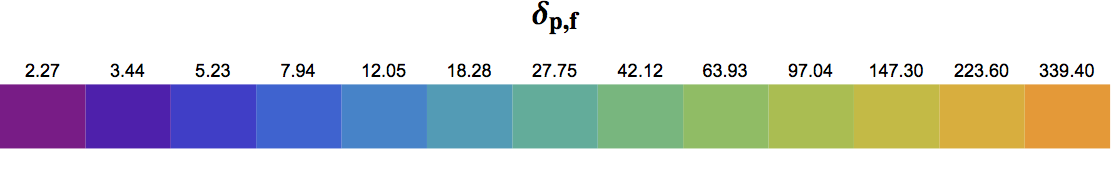}\\

\includegraphics[height=0.31\textwidth]{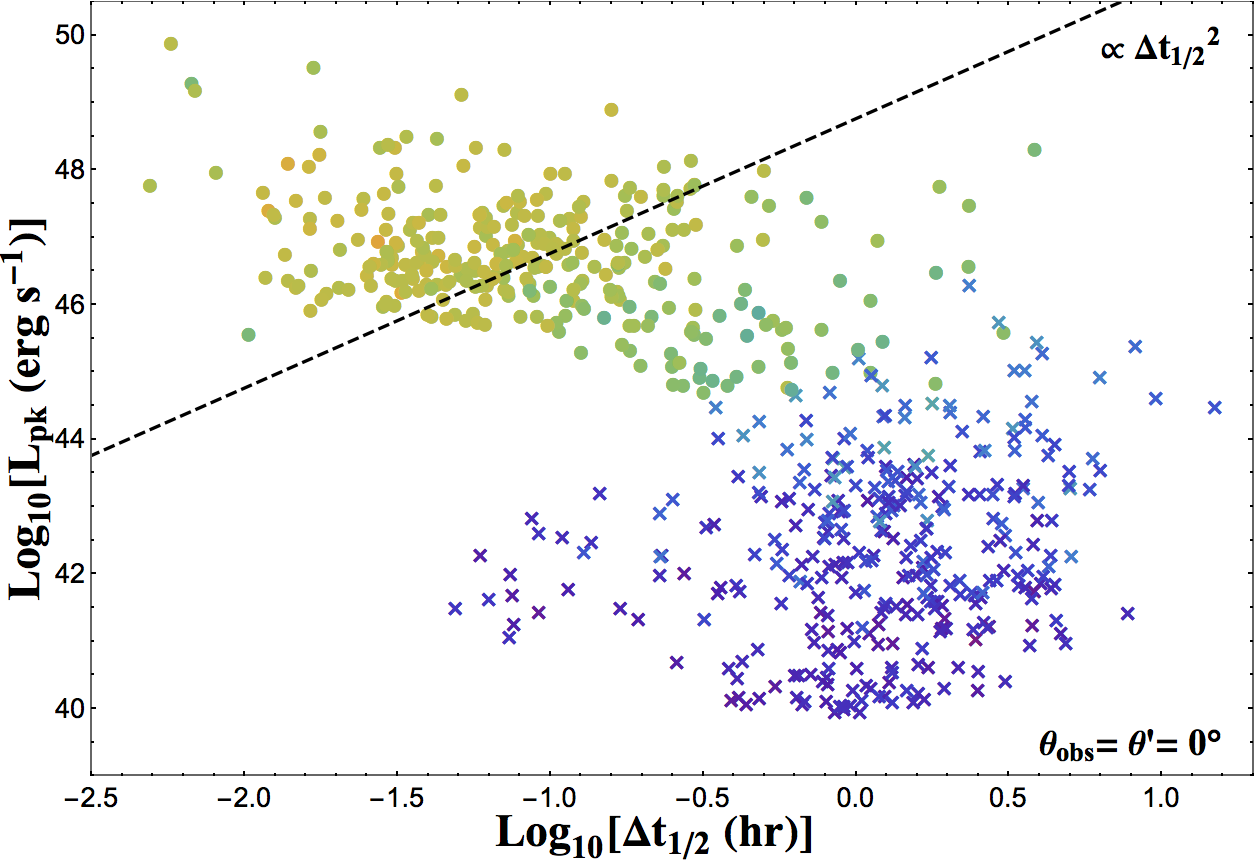}
\hspace{0.2cm}
\includegraphics[height=0.31\textwidth]{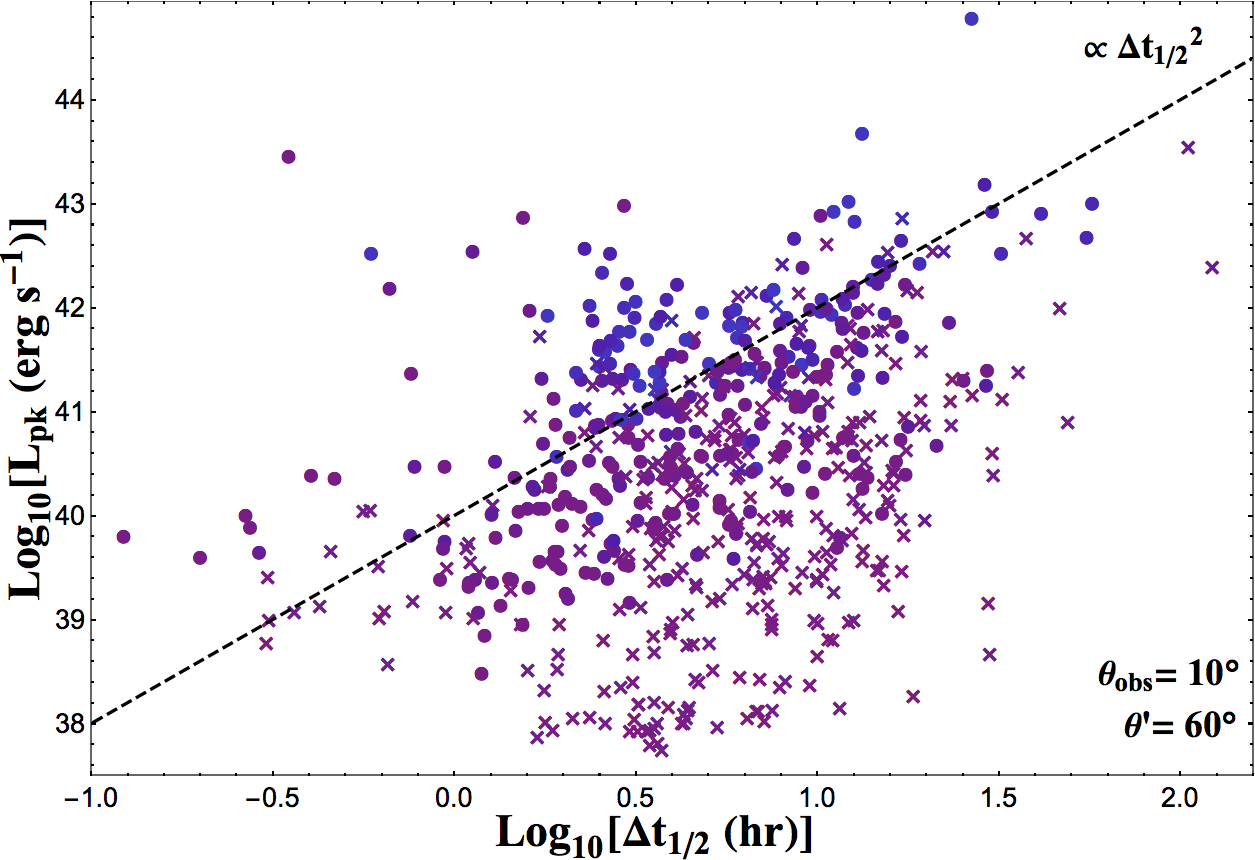}
\caption{Distributions of the peak bolometric luminosity $L_{\rm pk}$ and flux doubling timescales $\Delta t_{1/2}$, as produced from F10, as a function of the plasmoid's final transverse size, normalized to $L$, (top row) and Doppler factor (bottom row) (see legend above each row; a coloured version of this plot is available online). The left and right panels in each row denote perfect and non-perfect (i.e. $\theta_{\rm obs} = 10^{\circ}$ and $\theta^{\prime} = 60^{\circ}$) alignment, respectively. The plot-marker type corresponds to whether a plasmoid is moving on the side of the reconnection layer oriented towards or away from an observer, respectively (see Figs.~\ref{fig:tracks} and~\ref{fig:layer_in_jet_sketch}). The dashed lines denote scaling of $L_{\rm pk}$ with $\Delta t_{1/2}$ for either constant $w_{\perp, f}$ (top row) or constant $\delta_{\rm p, f}$ (bottom panel), as determined from the analytical estimates of PGS16.}
\label{fig:sig_10_flare_distribution_size_dependence}
\end{figure*}
%%%%%% FIGURE END %%%%%%%%

We can decompose the light curves presented in the previous section into the contributions of individual plasmoids. In particular, we compute the peak bolometric luminosities $L_{\rm pk}$ and flux doubling timescales\footnote{Defined as the time needed for the luminosity to increase from $L_{\rm pk}/2$ to $L_{\rm pk}$ in the observer's frame.} $\Delta t_{1/2}$ of all plasmoids. 
Our results for various orientations as produced for F10 are provided in Fig.~\ref{fig:sig_10_flare_distribution_size_dependence}. Here, we plot the distributions for $L_{\rm pk}$ and $\Delta t_{1/2}$ assuming perfect and non-perfect (i.e. $\theta_{\rm obs} = 10^\circ$ and $\theta^{\prime} = 60^\circ$) alignment in the left and right panel, respectively. The colour coding in the top and bottom rows denote a plasmoid's final transverse size, normalized to $L$, and the Doppler factor, respectively. The plot markers indicate whether a plasmoid's final location is on the side of the reconnection layer oriented towards or away from the observer (see Figs.~\ref{fig:tracks} and~\ref{fig:layer_in_jet_sketch}). 

In the left panels of Fig.~\ref{fig:sig_10_flare_distribution_size_dependence}, we find a clear separation of plasmoids on opposite sides of the reconnection layer. This particular orientation results in large Doppler boosting of small to medium sized plasmoids on the right side of the layer (see Fig.~\ref{fig:cdf_final_doppler_factor}) such that they have large peak bolometric luminosities and small flux-doubling timescales. The opposite is found for those plasmoids oriented away from an observer. The separation of plasmoids, based on location within the reconnection layer is less prominent in the right panels of Fig.~\ref{fig:sig_10_flare_distribution_size_dependence}. For this particular orientation, the majority of plasmoids have $\delta_{\rm p, f} \sim \delta_{\rm j} \sim 1$ resulting in low-Doppler boosting of all individual plasmoid-powered flares. Because of this, the observed luminosities are comparable to the co-moving luminosities of each plasmoid.

Overplotted in the top row of Fig.~\ref{fig:sig_10_flare_distribution_size_dependence} is a black, dashed line representing the power-law scaling $L_{\rm pk} \propto \Delta t_{1/2} ^{-4}$, which is valid for fixed plasmoid size (for a comparison, see Fig.~17 in \cite{pcsg17}). 
Our results exhibit this parametric scaling as the dashed lines intersect points of similar colour, which correspond to plasmoids of similar sizes. However, there are a few new features shown in Fig.~\ref{fig:sig_10_flare_distribution_size_dependence} that were not present in previous studies, namely the separation of plasmoids based on location within the reconnection layer (i.e. considered emission from both sides of the layer) and the production of more luminous flares, as compared to Fig.~17 in \cite{pcsg17}. These differences arise from the fact that PGS16 and \cite{pcsg17} considered the emission only from one side of the layer and did not include emission from ECS but only SSC.

%In Fig.~\ref{fig:sig_10_flare_distribution_doppler_dependence}, we show a similar scatter plot of the distributions of $L_{\rm pk}$ and $\Delta t_{1/2}$, for $\sigma = 10$ and perfect alignment, as a function of the plasmoids' final Doppler factor. As previously mentioned, those plasmoids oriented towards an observer can obtain large values of $\delta_{\rm p, f}$, while those oriented away have $\delta_{\rm p, f} \sim 1$. 
Overplotted  in the bottom row is a dashed line corresponding to the parametric scaling of $L_{\rm pk} \propto \Delta t_{1/2}^{2}$, valid for fixed $\delta_{\rm p}$ but varying $w_{\perp,f}$ (PGS16). We find this parametric scaling still valid as the line intersects points with the same colour, which correspond to plasmoids with similar Doppler factor. An estimate for the maximum allowed Doppler factor, in perfect alignment, is $\sim 4 \Gamma_{\rm j} \sqrt{1+\sigma}$ \citep{giannios2009}. However, we find that a few plasmoids obtain values larger than this, which is most likely due to bulk acceleration of a few plasmoids which are born in the vicinity of large plasmoids and are attracted to it (see Fig.~4 in \cite{pcsg17}).

Although the analytical estimates of PGS16 were derived assuming perfect alignment and considered plasmoids from only one side of the layer, it is interesting to find that they are valid for all orientations. The comparison provided in this section strengthens the validity of the authors' previous studies. This allows for a simplified application of the Monte Carlo results of \cite{pcsg17} to many reconnection events of different $\sigma$. Doing so provides an insight into the statistical properties of the reconnection event observables.

\section{Summary \& Discussion}
\label{sec:discussion}

Relativistic magnetic reconnection is a highly dynamical process which can naturally produce a chain of plasmoids, each containing relativistic particles and magnetic fields. The properties of these plasmoids, which are ideal candidates for the emission sites in blazars, can only be studied through kinetic PIC simulations. 
Here, we adopted the 2D PIC results of relativistic reconnection in pair plasmas by SGP16 and along with our radiative transfer model we computed the cumulative emission (spectra and light curves) from an entire reconnection event.

Although the adopted 2D PIC results of SGP16 are fully equipped to track the dynamics and evolution of all plasmoids and particles within the reconnection layer, we require a simplification of the plasmoid-model. In Sec.~\ref{sec:plas_select}, we made several assumptions which allowed for easier computation of the cumulative light curves from a single reconnection event. The most important of which assumes plasmoids are homogeneous structures. As is clearly shown in SGP16, plasmoids have radial dependence in virtually all of their properties, with an increase in the particle number density and magnetic field strength towards the plasmoid center. Let us consider here, a two-zone toy model composed of an inner region, defined from the center of the plasmoid to the radial distance at which its density reaches half of its maximum value $r_{\rm in}$, and an outer region, defined from the aforementioned distance to the outer edge of the plasmoid $r_{\rm out}$. For any given plasmoid, the ratio of $r_{\rm out} / r_{\rm in}$ can range from $\sim 1$ to $\sim 10$. %The values of $r_{\rm in}$ can range up to $\sim 0.001 \, L$ while $r_{\rm out}$ can range up to $\sim 0.05 \, L$. %\LS{would it make more sense to quote the range of the ratio for a GIVEN plasmoid?} 
From the parametric scalings provided in Appendix~A of SGP16, we know that the particle number density within a plasmoid goes as $\propto (y/r_{\rm out})^{-1}$ and the magnetic energy fraction goes as $\propto (y/r_{\rm out})^{-1.2}$, where $y$ is the radial distance from the plasmoid's center. By integrating over the two regions and assuming a similar particle distribution in both, we can estimate the ratio of the synchrotron power from the outer and inner regions, which scales as $(r_{\rm out}/r_{\rm in})^{4/5}$ for $r_{\rm out}\gg r_{\rm in}$. We can thus conclude that the outer region would dominate the synchrotron radiation output of the plasmoid. Similar estimates can be made for the Compton emission from the two regions. Because of this, we may assume a homogeneous one-zone model, in which the plasmoid's area averaged properties are approximately equal to those averaged over the outer region of the plasmoid. 
%\MP{The bottom line is missing. Do we miss a lot by using the averaged properties of a plasmoid?}

Additionally, in the context of a homogeneous one-zone model, we lose the ability to perform and predict any polarization signatures from the reconnection layer. Optical polarization monitoring of blazars have observed swings in the polarization angle which are often accompanied by multi-wavelength flares \citep{marscher2008,marscher2010}. Studies have suggested these polarization angle swings are the result of physical processes altering the magnetic fields surrounding the emission sites \citep{zhang2016}. Through 2D PIC simulations, it has been shown that the relativistic reconnection process can produced variable optical and X-ray polarization signatures \citep{zhang2018,tavecchio2018}. 

We have used a simplified description of particle acceleration in our radiative transfer code (see Sect.~\ref{sec:model}). For example, the energy spectrum of particles accelerated during mergers was taken to be the same as that of particles accelerated outside the plasmoids, while their injection was modeled as a smooth function of time. Instead, a merger could be modeled as an instantaneous episode of  injection of particles with a bias towards high energies. Although the number of particles accelerated during a merger may be small, as compared to those accelerated at X-points, they might result in unique observational signatures (e.g. flares with spectral hardening). %can reach large Lorentz factors which might result in unique observational signatures (e.g., flares with spectral hardening). 
%\LS{I am not 100 percent sure that this balance of number of accelerated particles is true, maybe we can soften the previous sentence (or just mention that it may result in hardening)} 
Moreover, particles could gain energy while residing in a plasmoid due to compression, as recently demonstrated by \cite{petropoulou2018}. We plan to expand our radiative transfer code to account for these effects as part of a future publication.

Our approach is one of the first to provide a physically motivated model for explaining the multi-timescale and multi-wavelength variability. 
Here, we showed that a light curve produced from a single reconnection event is composed of numerous powerful flares with timescales ranging on several orders of magnitude. 
A direct result of our model is the prediction of multi-wavelength flares, occurring at optical, UV, X-ray, and $\gamma$-ray  energies (see Fig.~\ref{fig:individual_spectra_plots_fsrq} and~\ref{fig:individual_spectra_plots_bl_lac}).
Thus, a promising strategy to compare our model results to multi-wavelength blazar observations would be to investigate the correlation of flaring events in different wavelength bands (Christie et al., in prep.). The radio variability observed in blazars \citep{richards2011}, however, can not be directly accounted for by our model due to the strong presence of synchrotron self-absorption while the plasmoids are within the reconnection layer. After plasmoids are advected out of the layer, the radio emission could become important as plasmoids may undergo adiabatic expansion and their particle distributions cool to lower energies, resulting in an optically thin radio spectrum.

In our modeling of FSRQ-like objects, the emitting regions are located within the BLR, thereby requiring the photons escaping from plasmoids to propagate through this region. We can compute the optical depth for photon-photon absorption of $\gamma$-rays on BLR photons using the usual expressions \citep{coppi1990} for a blackbody emitter with characteristic frequency $\sim$few eV. We find that the attenuation becomes important for $\gamma$-rays at energies $\gtrsim 600$~GeV. However, this is not a significant effect within our model as the max frequencies reached for F3 and F10 are $\sim 300$~GeV (see Fig.~\ref{fig:individual_spectra_plots_fsrq}). Additionally, our choice of $R_{\rm BLR}\gtrsim z_{\rm diss}=5\times10^{17}$~cm, implies a very luminous accretion disk, which could be present in galaxies with black holes of mass $\sim 10^9 M_{\odot}$. Modeling FSRQ-like objects can also be achieved with a smaller disk luminosity while assuming a smaller value of $L$ and $z_{\rm diss}$. Although lowering $L$ decreases the bolometric luminosity, a similar Compton ratio ($\propto \Gamma_{\rm p}^{2} U_{\rm BLR}/U_{\rm B}$) can be obtained by slightly lowering the magnetic field strength of plasmoids. For fixed $\sigma$ and $L$ (e.g. F10 and B10), one can obtained different luminosity flares by varying two free parameters. %\MP{What about L? We discussed above, but maybe you want that in addition to L, we can change the luminosity by ...} 
The first is $L_{\rm BLR}$: increasing its value results in a potentially larger Compton dominance, assuming the dissipation region falls within the BLR. The second is the magnetic field strength $B$: increasing its values would yield a higher luminosity synchrotron component while simultaneously lowering the SSC component. Higher magnetic field strengths would also result in shorter cooling timescales of the injected particles, thereby steepening the decay portion of a plasmoid's light curve. %\MP{Maybe here, you can append the paragraph where we discuss different model parameters?}

%\MP{It would be useful for readers is to mention explicitly how you get different luminosities in the model for fixed sigma. What do we need to adjust? Mention this clearly somewhere.} \MP{This paragraph could be placed in the discussion section}
%\textbf{For fixed $\sigma$, such as F10 and B10, one can obtain different luminosity flares by varying two free parameters. The first is $L_{\rm BLR}$: increasing its value results in a potentially larger Compton dominance, assuming the dissipation region falls within the BLR. The second is the magnetic field strength $B$. The authors in PGS16 showed using analytical estimates that the bolometric luminosity of an individual plasmoid flare scales as $\propto B^{2}$. \MP{Where do we say that exactly?} Although increasing $B$ would result in higher luminosity flares, particles would also cool rapidly, resulting in a steeper decay.}

Current sheets moving with the bulk flow of the jet would eventually reach and traverse the assumed BLR (which is relevant for the FSRQ-like models) on an observed timescale of $t \approx R_{\rm BLR}/(c \Gamma_{\rm j} \delta_{\rm j}) \sim$~ a few days. Thus, any emission produced beyond this time would have a much lower Compton emission than that presented in Fig.~\ref{fig:gamma_ray_lc_fsrq}. %Thus, any emission produced beyond this time would have a much lower Compton emission than the one shown in the $\gamma$-ray light curves presented in Fig.~\ref{fig:gamma_ray_lc_fsrq} would display a large drop in luminosity. %\LS{sth wrong with previous sentence} 
Whether current sheets move with the bulk flow or are stationary features depends on their formation mechanism (e.g., striped wind or kink instability) \citep{giannios2006,parfrey2015,barniol2017,giannios2018}.

Turbulence within the bulk flow of the jet can also play an important role on the observed duration and stability of a reconnection event. If the layer was formed due to kink instabilities, turbulent eddies within the bulk flow could disrupt the reconnection process on a timescale of $\sim \varpi /c$, where $\varpi$ is the cross-section of the jet, thereby limiting the observed duration of the event. However, if we considered the striped wind model as the formation mechanism, there would be pre-existing current sheets within the bulk flow and turbulence would not be expected, except possibly at the outer edge of the dissipation zone \citep{zrake2017}. Additionally, the structure of current sheets within a global jet model is an interesting topic which has not been properly addressed. By use of global MHD simulations, one would be able to assess the statistical properties of current sheets (e.g. length, magnetization) and distinguish which case (i.e. kink instability or striped-wind) is most relevant for observations.

%\MP{This paragraph appears disconnected , but I do not have a better idea about where to place it!} \textbf{I agree but I don't know where to put it.}
A strength of our blazar emission model is that its dynamical evolution is directly dictated by the PIC simulations. Because of this, we have a small number of free parameters; namely, the magnetization $\sigma$, pair multiplicity $N_\pm$, magnetic field strength $B$, bulk Lorentz factor of the jet $\Gamma_{\rm j}$, orientation angles $\theta^{\prime}$ and $\theta_{\rm obs}$, and half-length of the reconnection layer $L$. The first two parameters set the slope and range of the injected particle distribution and its corresponding SED. The remaining four parameters are key in determining the luminosity and duration of all flares. We found that the observed SEDs of BL Lacs and FSRQs favor dissipation regions with $\sigma > 10$ for the former and $\sigma < 10$ for the latter. Interestingly, the transition from a BL Lac to an FSRQ can be achieved by taking into account the BLR photons and assuming an electron-proton jet for the former and  pair-rich jet (with several pairs per proton) in the latter, while all remaining model parameters can remain fixed.

\section{Conclusion}
This study is a step towards solidifying the relativistic magnetic reconnection as the process responsible for the production of the multi-wavelength spectral and temporal variability observed in blazars. A comparison of our model predictions with observations will allow us to constrain properties of the emission sites within blazar jets.
% \MP{Maybe find a catchy concluding sentence from our fermi proposal?}

% Additional \MP{steps for verifying this model} include the comparison of the distributions of flares (luminosities and durations) from a reconnection event with those obtained from blazar observations as well as correlation of flares occurring in different wavelength bands. 

%\MP{I am not sure that the last paragraph fits here. Maybe you want to close with the main conclusion of this work and an outlook. }
%Although we adopted the 2D PIC results of SGP16 for an application to blazar flaring events, it it not limited to this specific astrophysical phenomena. Relativistic magnetic reconnection is believed to play a dominant role in the prompt emission of gamma-ray burst (GRBs) \citep{lyutikov2003,giannios2008,beniamini2018}, pulsar wind nebulae (PWNe) \citep{cerutti2016,cerutti2017}, and flaring events from accreting black holes \citep{beloborodov2017}. A similar application of the 2D PIC results would provide a useful insight into the observed high-energy emission of these various phenomena.

\section*{Acknowledgements} 

%We thank XXX for useful discussion and comments.    
We thank the referee for his/her constructive report that helped to improve the manuscript. We also thank Drs. Q.~Feng, J.~Finke, and T.~Hovatta for useful comments. I.C. and D.G. acknowledge support from NASA ATP grants NNX16AB32G and NNX17AG21G. L.S. acknowledges support from DoE DE-SC0016542, NASA Fermi NNX-16AR75G, NASA ATP NNX-17AG21G, NSF ACI-1657507, and NSF AST1716567.
M.P. acknowledges support from the Lyman Jr.~Spitzer Postdoctoral Fellowship.
%%%%%%%%%%%%%%%%%%%%%%%%%%%%%%%%%%%%%%%%%%%%%%%%%%%%%%%%%%%%%%%%%%%%%%%%%%%%%%%
%%%%%%%%%%%%%%%%%%%%%%%%%%%%%%%%%%%%%%%%%%%%%%%%%%%%%%%%%%%%%%%%%%%%%%%%%%%%%%%

\bibliographystyle{mnras} % style mn2e.bst
\bibliography{bib.bib} % your references Yourfile.bib

%%%%%%%%%%%%%%%%%%%%%%%%%%%%%%%%%%%%%%%%%%%%%%%%%%%%%%%%%%%%%%%%%%%%%%%%%%%%%%%
%%%%%%%%%%%%%%%%%%%%%%%%%%%%%%%%%%%%%%%%%%%%%%%%%%%%%%%%%%%%%%%%%%%%%%%%%%%%%%%
\appendix
\section{Injected Particle Distribution}
\label{sec:particle_distribution}
The range of the injected particle distribution, i.e., $\gamma_{\rm min}$ and $\gamma_{\rm max}$, is partially determined through PIC. To determine these values, we start from equipartition between the relativistic pairs and magnetic fields within in a plasmoid, 
\eqb
\label{eqn:equipartition}
N_{\pm} n_{\rm co} m_{\rm e} c^{2} (\bar{\gamma} - 1) \approx \frac{B^2}{8 \pi},
\eqe
where $\bar{\gamma}$ is the characteristic Lorentz factor of the injected distribution, $n_{\rm co}$ is the time averaged co-moving particle number density determined using eqn.~\ref{eqn:physical_number_density}, and $N_{\pm}$ is the pair multiplicity. The latter variable is still unknown for blazar jets and we treat it as a free parameter. %and show its relevance to the photon spectra produced by an individual plasmoid (see Sec.~4.1).

For a distribution with slope $p>2$, the characteristic Lorentz factor, assuming $\gamma_{\rm max} \gg \gamma_{\rm min}$, is
\eqb
\label{eqn:gamma_char_1}
\bar{\gamma} \approx \frac{1 - p}{2 - p} \gamma_{\rm min}.
\eqe
For $p<2$, $\bar{\gamma}$ becomes
\eqb
\label{eqn:gamma_char_2}
\bar{\gamma} \approx \frac{p - 1}{2 - p} \gamma_{\rm min}^{p - 1} \gamma_{\rm max}^{2 - p}.
\eqe

For $\sigma \leq 10$, the distribution's slope is $p >2$, corresponding to a majority of the particle's energy being held at $\gamma_{\rm min}$. Using eqns.~\ref{eqn:equipartition} and~\ref{eqn:gamma_char_1}, we can estimate $\gamma_{\rm min}$ as
\eqb
\label{eqn:g_min}
\gamma_{\rm min} \approx \frac{p-2}{p-1} \left(\frac{4 \sigma m_{\rm p}}{n_{\rm PIC} \, m_{\rm e} N_{\pm}} + 1\right).
\eqe
For $\sigma \gg 10$, $p < 2$, indicating the energy is held by particles at $\gamma_{\rm max}$. Using eqns.~\ref{eqn:equipartition} and~\ref{eqn:gamma_char_2}, we obtain
\eqb
\label{eqn:g_max}
\gamma_{\rm max} \approx \left[\frac{2 - p}{p - 1} \, \gamma_{\rm min}^{1-p} \, \left(\frac{4 \sigma m_{\rm p}}{n_{\rm PIC} \, m_{\rm e} N_{\pm}} + 1\right)\right]^{1/(2 - p)}.
\eqe
%%%%%% FIGURE BEGIN %%%%%%%%
\begin{figure}
\centering
\includegraphics[height=0.3\textwidth]{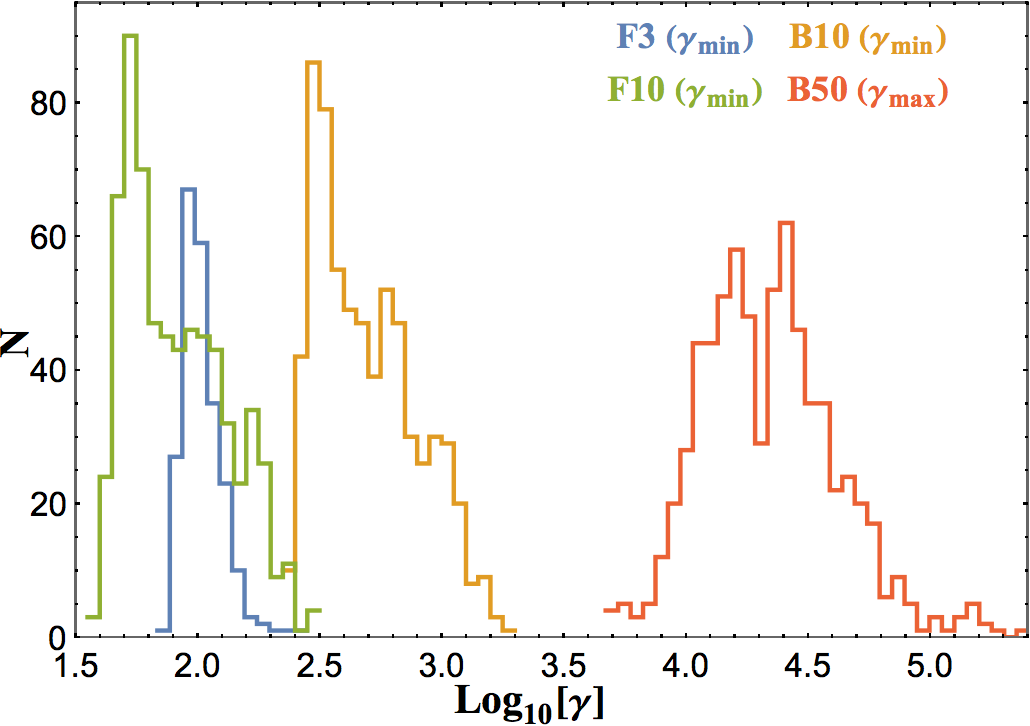}
\caption{Histogram displaying the distribution of the characteristic Lorentz factors of the injected particle distribution (see Appendix~\ref{sec:particle_distribution}) for the different $\sigma$ values listed on the plot. The values of $\gamma_{\rm min}$ and $\gamma_{\rm max}$ are determined using the parameters listed in Table~\ref{table:model_parameters_2} and eqns.~\ref{eqn:g_min} and \ref{eqn:g_max}, respectively. A coloured version of this plot is available online.}
\label{fig:particle_Lorentz_factor}
\end{figure}
%%%%%% FIGURE END %%%%%%%%

Because the values of $n_{\rm co}$ and $n_{\rm PIC}$ change for each plasmoid, there will be a range of $\gamma_{\rm min}$ and $\gamma_{\rm max}$ for each $\sigma$ value. Histograms of $\gamma_{\min}$ and $\gamma_{\max}$ for all plasmoids in each PIC simulation are presented in Fig.~\ref{fig:particle_Lorentz_factor}. The increasing spread found in the distributions for larger $\sigma$ are a direct result of the increasing spread in the particle number density per plasmoid (see panels d-f in Fig.~5 of SGP16).
However, for a particular distribution, we require both $\gamma_{\rm min}$ and $\gamma_{\rm max}$. For the parameter which is not determined by the estimates provided above is a free parameter within our model and is prescribed manually. We therefore, choose appropriate values (see Table~\ref{table:model_parameters_2}) such that the spectra of all plasmoids are similar to those observed in both BL Lacs and FSRQs.

\section{Doppler Factor \& Observer Time}
\label{sec:observer_time}
%%%%%% FIGURE BEGIN %%%%%%%%
\begin{figure*}
\centering
\includegraphics[width=0.8\textwidth]{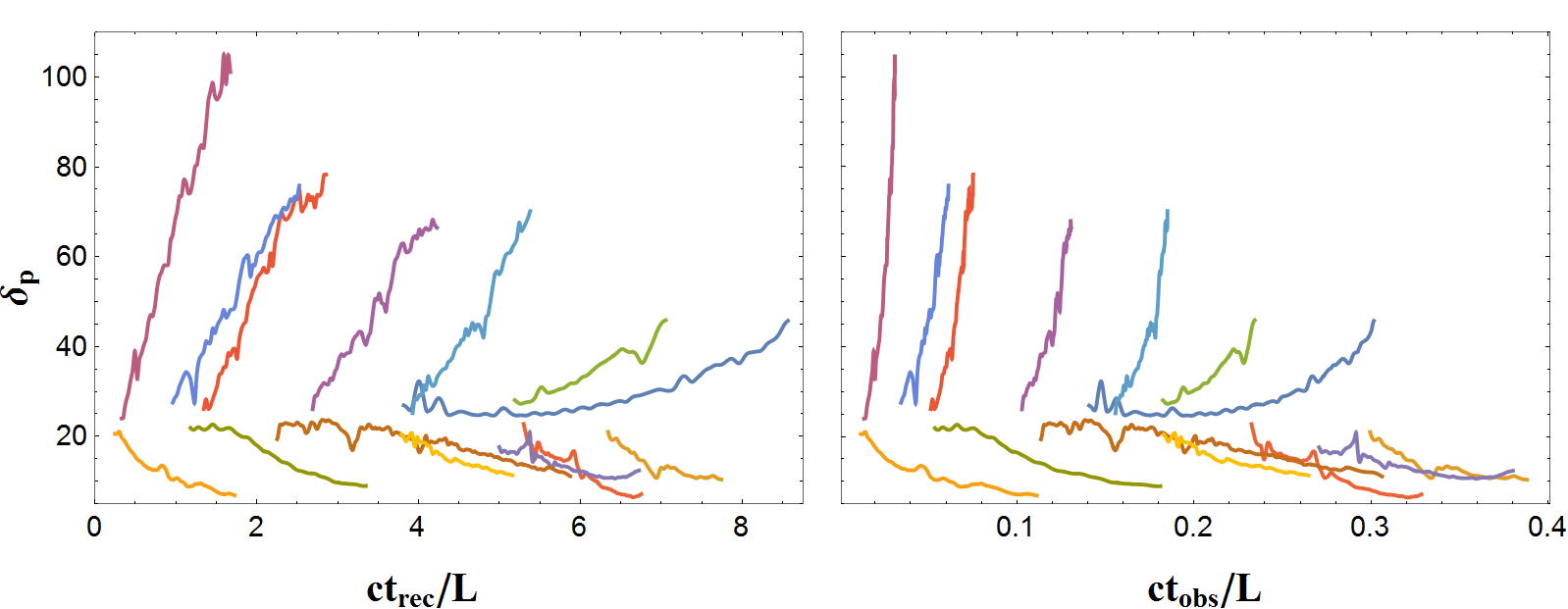}
\caption{The Doppler factor $\delta_{\rm p}$, as measured by an observer, as produced from a $\sigma = 10$ PIC simulation for 14 long-lived plasmoids as a function of the reconnection time (Left Panel) and the observer time (Right Panel), both of which are normalized to $L/c$. Note that the curves are produced using the parameters listed in Table~\ref{table:model_parameters_2}, namely $\Gamma_{\rm j} = 12$, and correspond to perfect alignment (i.e. $\theta_{\rm obs} = \theta^{\prime} = 0^{\circ}$), and are plotted up to moment of the plasmoid's death either by merger or advection from the layer. A coloured version of this plot is available online.}
\label{fig:doppler_factor_t_obs}
\end{figure*}
%%%%%% FIGURE END %%%%%%%%
As described in Sections~\ref{sec:pic_results} and~\ref{sec:results}, we numerically solve for the particle and photon distributions in the co-moving frame of each individual plasmoid. Once completed, we are then required to map all results to that seen by an observer, positioned at angle $\theta_{\rm obs}$ relative to the jet axis (see Figure~\ref{fig:layer_in_jet_sketch}). To complete this, the Doppler factor $\delta_{\rm p}$ of a plasmoid and the observer time of a flare $t_{\rm obs}$ must be known. The former is defined in eqn.~(\ref{eqn:Doppler_factor}), while the latter is determined as:

\eqb
\label{eqn:t_obs_1}
t_{\rm obs} (t) = t_{0} + \int_{t_{i}}^{t} \, \frac{{\rm d}t^{\prime}}{\delta_{\rm p}(t^{\prime}) \, \Gamma(t^{\prime})},
\eqe
where $t_{i}$ is the time, as measured in the reconnection frame, in which a plasmoid is born and $t_{0} = t_{i} / \delta_{\rm p}(t_i) \, \Gamma(t_i)$. As discussed in Sec.~\ref{sec:radiation_transfer}, we continuously solve the transport equations for the particle and photon distributions for a fraction of a dynamical time after a plasmoid has either advected from the layer or merged with a neighboring plasmoid. The time an observer measures post advection or merger is then
\eqb
\label{eqn:t_obs_2}
t_{\rm obs} (t) = t_0 + \int_{t_{i}}^{t_f} \, \frac{{\rm d}t^{\prime}}{\delta_{\rm p}(t^{\prime}) \, \Gamma(t^{\prime})} + \frac{t - t_f}{\delta_{\rm p}(t_f) \, \Gamma(t_f)},
\eqe
where $t_f$ is the time, as measured in the reconnection frame, in which a plasmoid mergers or advects. We note that when solving the transport equations post merger or advection, we assume the plasmoid is moving with the $\Gamma$ value just before it died.

In the left and right panels of Fig.~\ref{fig:doppler_factor_t_obs}, we plot the observed Doppler factor, for the 14 longest living plasmoids in the $\sigma=10$ PIC simulation, as a function of the reconnection and observer's time, respectively. Both times are normalized to $L/c$ for reference. These curves, which are displayed for perfect alignment, show both a rise and decline in $\delta_{\rm p}$ which correspond to a plasmoid moving towards or away from an observer, respectively. The variations seen in $\delta_{\rm p}$ are directly produced from variations in the plasmoids' Lorentz factor $\Gamma$ and can confidently be kept as PIC properly tracks the plasmoids' motion. 

\section{Smoothing PIC Results}
\label{sec:smoothing}
%%%%%% FIGURE BEGIN %%%%%%%%
\begin{figure}
\centering
\includegraphics[width=0.45\textwidth]{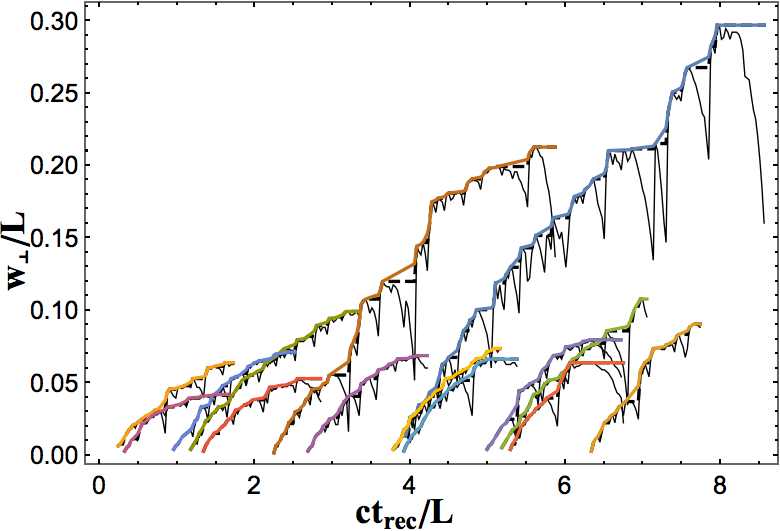}
\caption{Temporal evolution of the transverse size $w_{\perp}/L$ as a function of the reconnection time, in units of $L/c$, for the 14 longest living plasmoids as produced from a $\sigma=10$ PIC simulation. Each line style denotes a raw or manipulation of the PIC data. The thin, black lines represent the raw PIC data. The dashed black lines denote our first step in manipulation in which we apply the criteria that the plasmoid's size never be a decreasing function. The coloured lines denote the second step in our manipulation of $w_{\perp}$, in which all plateaus, seen in the dashed black lines, are connected linearly. A coloured version of this plot is available online.}
\label{fig:transverse_size_maipulation}
\end{figure}
%%%%%% FIGURE END %%%%%%%%

The PIC data, which described the plasmoids' properties and are used as the inputs for our radiative transfer model, are not smooth functions of time. In this section we describe the smoothing performed for each plasmoid property used within our model.

We begin with the plasmoid's transverse size $w_{\perp}$. As shown by the thin, black lines in Fig.~\ref{fig:transverse_size_maipulation} for the 14 longest living plasmoids as produced from a $\sigma = 10$ simulation, the raw data is found to contain large drops in $w_{\perp}$. This sharp decrease of the transverse size is a numerical artifact of the method used to track plasmoids during a merger, based on contours of the vector potential. To remove these drops, we apply a criterion which is preformed in two steps. The first is to state that the size is a monotonically increasing function, thereby giving $w_{\perp}$ a "staircase'' shape, as shown by the dashed, black lines in Fig.~\ref{fig:transverse_size_maipulation}. This introduces plateaus where drops previously existed. The second step is to then remove the plateaus by replacing them with a linear fit as to avoid introducing a vanishing injection rate, as we will  discuss below. The results for $w_{\perp}$ used within our model (e.g. co-moving volume and the escape time) are shown by the individual coloured lines in Fig.~\ref{fig:transverse_size_maipulation}.
    
The co-moving volume of a plasmoid is estimated by using $w_{\perp}(t)$, which is determined through PIC. However, the transverse size of a plasmoid is not a smooth function and will lead to variations in the volume, $Q_{\rm inj}^{\rm e} (\gamma, t)$, and therefore the luminosity produced by an individual plasmoid. An example of these variations by using the raw data from PIC is shown by the blue curves in the top and bottom panels of Fig.~\ref{fig:smoothing_test}, which displays $\partial_t (V(t)/L^3)$ and the bolometric, co-moving luminosity for a large plasmoid of final size $w_{\perp, f}/L \sim 0.3$ as produced from a $\sigma=10$ simulation. 

The second quantity which requires manipulation is ultimately related to the plasmoid's co-moving volume $V_{\rm co}(t)$, namely the instantaneous particle injection rate $Q_{\rm inj}^{\rm e}$ (see eqn.~\ref{eqn:q_inj}). This quantity is determined by taking the time derivative of $V(t)$, or more specifically the transverse size $w_\perp$. Although $w_\perp$ is always an increasing function, it is not a generally smooth function, thus leading to large variations in its time derivative. This can be seen by the blue curve in the top panel of Fig.~\ref{fig:smoothing_test} and explains our reasoning in removing the plateaus in $w_\perp$ (see black, dashed lines in Fig.~\ref{fig:transverse_size_maipulation}). These larges variations in $\partial_{\rm t} w_\perp$ and $Q_{\rm inj}$ result in variations in the co-moving light curve of an individual plasmoid (see blue curve in bottom panel of Fig.~\ref{fig:smoothing_test}). It should be noted that these variations are unphysical as they result from our manipulation of $w_\perp$ and not, for example, from variations in the plasmoid's Lorentz factor $\Gamma$. To remove these variations from the light curve, we apply a log-based Gaussian filter to $Q_{\rm inj}^{\rm e}$, whose range is dependent upon the final transverse size of the plasmoid, as the larger size relates to a long-lived plasmoid. 

To investigate the role of the filter's range on the light curve, we plot two examples, along with the raw data, in Fig.~\ref{fig:smoothing_test}. The orange curves displays a smoothing of $Q_{\rm inj}^{\rm e}$ such that all sharp features are removed from the raw data (top panel) and the resulting light curve (bottom panel). Although sharp features are removed, we still find large variations in the bolometric light curve. The green curves show the results for our sized-based smoothing method in which all variations are removed, resulting in a smooth increasing light curve. This method is performed for all plasmoids within our model thereby allowing all variations in the observed light curves to originate from $\Gamma$ and the Doppler factor (see Fig.~\ref{fig:doppler_factor_t_obs}).

%%%%%% FIGURE BEGIN %%%%%%%%
\begin{figure}
\centering
\includegraphics[height=0.5\textwidth]{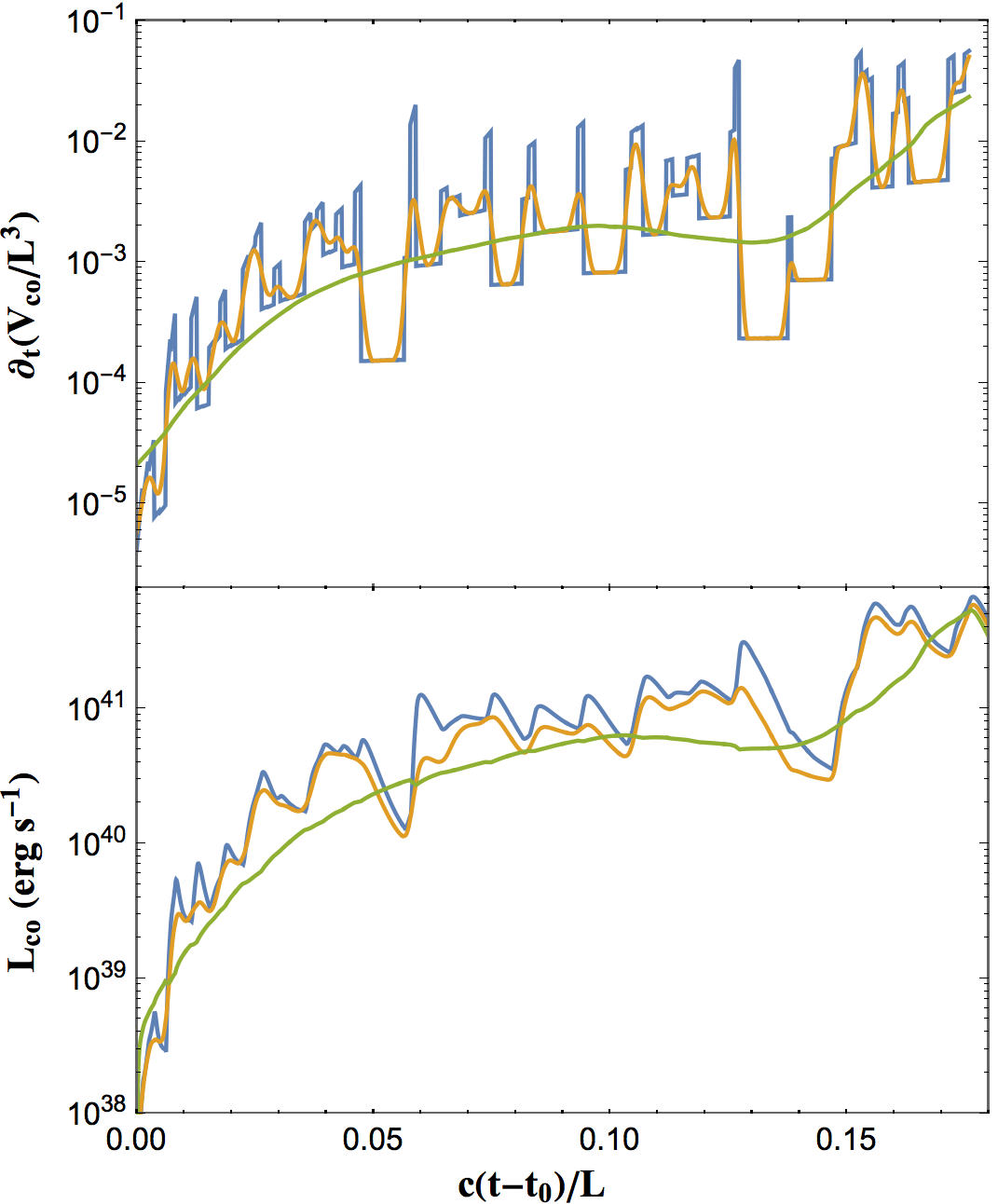}
\caption{Top Panel: Derivative of the co-moving volume, normalized to $L^3$, used to determine the instantaneous injection rate of particles, as a function of the plasmoid's lifetime, normalized to $L/c$. The blue curve denotes the raw data while the orange and green curves represent the raw data after different methods of smoothing  (a coloured version of this plot is available online). The sharp features occurring in the blue curve are due to our manipulation of the PIC data, specifically the plasmoid's transverse size $w_{\perp}$. Bottom Panel: Bolometric luminosity, as measured in the co-moving frame of the plasmoid, for the three cases of smoothing. The choice of smoothing numerical artifacts from our manipulation of the 2D PIC results greatly affects the resulting light curve and can produce unrealistic, synthetic ultra-fast flares on top of the individual plasmoid-powered flare. The results displayed here are produced for a \textit{monster} plasmoid ($w_{\perp, f} \sim 0.3 L$) from a $\sigma=10$ PIC simulation.}
\label{fig:smoothing_test}
\end{figure}
%%%%%% FIGURE END %%%%%%%%

\end{document}